\input harvmac
\input epsf
\overfullrule=0pt
\font\bigcmsy=cmsy10 scaled 1500
\def\Title#1#2{\rightline{#1}\ifx\answ\bigans
\nopagenumbers\pageno0\vskip1in
\else\pageno1\vskip.8in\fi \centerline{\titlefont #2}\vskip .5in}
\newcount\figno
\figno=0
\def\fig#1#2#3{
\par\begingroup\parindent=0pt\leftskip=1cm\rightskip=1cm\parindent=0pt
\baselineskip=11pt
\global\advance\figno by 1
\midinsert
\epsfxsize=#3
\centerline{\epsfbox{#2}}
\vskip 12pt 
{\bf Figure\ \the\figno: } #1\par
\endinsert\endgroup\par
}
\def\figlabel#1{\xdef#1{\the\figno}}
\def\encadremath#1{\vbox{\hrule\hbox{\vrule\kern8pt\vbox{\kern8pt
\hbox{$\displaystyle #1$}\kern8pt}
\kern8pt\vrule}\hrule}}
%\draftmode

\def \Bsw{\!\mathrel{\hbox{\bigcmsy\char'056}}\!}
\def \Bse{\!\mathrel{\hbox{\bigcmsy\char'046}}\!}
\def\lphi{{\Phi}}
\def\bphi{{\phi}}
\def\Boxx{{\partial_x^2}}
\def\Boxy{{\partial_y^2}}
\def\con{{\cal E}}

\def\gop{{\varphi}}

%\CecottiME
\lref\CecottiME{
  S.~Cecotti and C.~Vafa,
  ``Topological antitopological fusion,''
  Nucl.\ Phys.\  B {\bf 367}, 359 (1991).       
  %%CITATION = NUPHA,B367,359;%%
}

%\SeibergRS
\lref\SeibergRS{
  N.~Seiberg and E.~Witten,
  ``Monopole Condensation, And Confinement In N=2 Supersymmetric Yang-Mills
  Theory,''
  Nucl.\ Phys.\  B {\bf 426}, 19 (1994)
  [Erratum-ibid.\  B {\bf 430}, 485 (1994)]
  [arXiv:hep-th/9407087].
  %%CITATION = NUPHA,B426,19;%%
}

%\SeibergAJ
\lref\SeibergAJ{
  N.~Seiberg and E.~Witten,
  ``Monopoles, duality and chiral symmetry breaking in N=2 supersymmetric
  QCD,''
  Nucl.\ Phys.\  B {\bf 431}, 484 (1994)
  [arXiv:hep-th/9408099].
  %%CITATION = NUPHA,B431,484;%%
}

%\AharonyXZ
\lref\AharonyXZ{
  O.~Aharony, A.~Fayyazuddin and J.~M.~Maldacena,
  ``The large N limit of N = 2,1 field theories from three-branes in
  F-theory,''
  JHEP {\bf 9807}, 013 (1998)
  [arXiv:hep-th/9806159].
  %%CITATION = JHEPA,9807,013;%%
}

%\ArgyresWT
\lref\ArgyresWT{
  P.~C.~Argyres, M.~R.~Plesser and A.~D.~Shapere,
  %``The Coulomb phase of N=2 supersymmetric QCD,''
  Phys.\ Rev.\ Lett.\  {\bf 75}, 1699 (1995)
  [arXiv:hep-th/9505100].
  %%CITATION = PRLTA,75,1699;%%
}

%\EguchiVU
\lref\EguchiVU{
  T.~Eguchi, K.~Hori, K.~Ito and S.~K.~Yang,
  ``Study of $N=2$ Superconformal Field Theories in $4$ Dimensions,''
  Nucl.\ Phys.\  B {\bf 471}, 430 (1996)
  [arXiv:hep-th/9603002].
  %%CITATION = NUPHA,B471,430;%%
}

%\EguchiDS
\lref\EguchiDS{
  T.~Eguchi and K.~Hori,
  ``N = 2 superconformal field theories in 4 dimensions and A-D-E
  classification,''
  arXiv:hep-th/9607125.
  %%CITATION = HEP-TH/9607125;%%
}

%\MinahanFG
\lref\MinahanFG{
  J.~A.~Minahan and D.~Nemeschansky,
  ``An N = 2 superconformal fixed point with E(6) global symmetry,''
  Nucl.\ Phys.\  B {\bf 482}, 142 (1996)
  [arXiv:hep-th/9608047].
  %%CITATION = NUPHA,B482,142;%%
}

%\MinahanCJ
\lref\MinahanCJ{
  J.~A.~Minahan and D.~Nemeschansky,
  ``Superconformal fixed points with E(n) global symmetry,''
  Nucl.\ Phys.\  B {\bf 489}, 24 (1997)
  [arXiv:hep-th/9610076].
  %%CITATION = NUPHA,B489,24;%%
}

%\ArgyresPP
\lref\ArgyresPP{
  P.~C.~Argyres, M.~Crescimanno, A.~D.~Shapere and J.~R.~Wittig,
  `Classification of N = 2 superconformal field theories with  two-dimensional
  Coulomb branches,''
  arXiv:hep-th/0504070.
  %%CITATION = HEP-TH/0504070;%%
}

%\DixonFJ
\lref\DixonFJ{
  L.~J.~Dixon, V.~Kaplunovsky and J.~Louis,
  ``On Effective Field Theories Describing (2,2) Vacua of the Heterotic
  String,''
  Nucl.\ Phys.\  B {\bf 329}, 27 (1990).
  %%CITATION = NUPHA,B329,27;%%
}

%\WilczekDH
\lref\WilczekDH{
  F.~Wilczek and A.~Zee,
  ``Appearance Of Gauge Structure In Simple Dynamical Systems,''
  Phys.\ Rev.\ Lett.\  {\bf 52}, 2111 (1984).
  %%CITATION = PRLTA,52,2111;%%
}

%\IntriligatorJR
\lref\IntriligatorJR{
  K.~A.~Intriligator, R.~G.~Leigh and N.~Seiberg,
  ``Exact superpotentials in four-dimensions,''
  Phys.\ Rev.\  D {\bf 50}, 1092 (1994)
  [arXiv:hep-th/9403198].
  %%CITATION = PHRVA,D50,1092;%%
}

%\ArgyresWX
\lref\ArgyresWX{
  P.~C.~Argyres and J.~R.~Wittig,
  ``Classification of N = 2 superconformal field theories with  two-dimensional
  Coulomb branches. II,''
  arXiv:hep-th/0510226.
  %%CITATION = HEP-TH/0510226;%%
}

%\BerryJV
\lref\BerryJV{
  M.~V.~Berry,
  ``Quantal phase factors accompanying adiabatic changes,''
  Proc.\ Roy.\ Soc.\ Lond.\  A {\bf 392}, 45 (1984).
  %%CITATION = PRSLA,A392,45;%%
}

%\ArgyresTQ
\lref\ArgyresTQ{
  P.~C.~Argyres and J.~R.~Wittig,
  ``Infinite coupling duals of N=2 gauge theories and new rank 1 superconformal
  field theories,''
  JHEP {\bf 0801}, 074 (2008)
  [arXiv:0712.2028 [hep-th]].
  %%CITATION = JHEPA,0801,074;%%
}

%\GaiottoWE
\lref\GaiottoWE{
  D.~Gaiotto,
  ``N=2 dualities,''
  arXiv:0904.2715 [hep-th].
  %%CITATION = ARXIV:0904.2715;%%
}

%\KuzenkoPI
\lref\KuzenkoPI{
  S.~M.~Kuzenko and S.~Theisen,
  ``Correlation functions of conserved currents in N = 2 superconformal
  theory,''
  Class.\ Quant.\ Grav.\  {\bf 17}, 665 (2000)
  [arXiv:hep-th/9907107].
  %%CITATION = CQGRD,17,665;%%
}

%\deWitPK
\lref\deWitPK{
  B.~de Wit and A.~Van Proeyen,
  ``Potentials And Symmetries Of General Gauged N=2 Supergravity: Yang-Mills
  Models,''
  Nucl.\ Phys.\  B {\bf 245}, 89 (1984).
  %%CITATION = NUPHA,B245,89;%%
}

%\PetkouFV
\lref\PetkouFV{
  A.~Petkou and K.~Skenderis,
  ``A non-renormalization theorem for conformal anomalies,''
  Nucl.\ Phys.\  B {\bf 561}, 100 (1999)
  [arXiv:hep-th/9906030].
  %%CITATION = NUPHA,B561,100;%%
}

\lref\DixonFJ{
  L.~J.~Dixon, V.~Kaplunovsky and J.~Louis,
  ``On Effective Field Theories Describing (2,2) Vacua of the Heterotic
  String,''
  Nucl.\ Phys.\  B {\bf 329}, 27 (1990).
  %%CITATION = NUPHA,B329,27;%%
}

%\ArgyresBN
\lref\ArgyresBN{
  P.~C.~Argyres and A.~Buchel,
  ``The nonperturbative gauge coupling of N = 2 supersymmetric theories,''
  Phys.\ Lett.\  B {\bf 442}, 180 (1998)
  [arXiv:hep-th/9806234].
  %%CITATION = PHLTA,B442,180;%%
}

%\GreeneVM
\lref\GreeneVM{
  B.~R.~Greene, D.~R.~Morrison and M.~R.~Plesser,
  ``Mirror manifolds in higher dimension,''
  Commun.\ Math.\ Phys.\  {\bf 173}, 559 (1995)
  [arXiv:hep-th/9402119].
  %%CITATION = CMPHA,173,559;%%
}

%\DolanTT
\lref\DolanTT{
  F.~A.~Dolan and H.~Osborn,
  ``Superconformal symmetry, correlation functions and the operator product
  expansion,''
  Nucl.\ Phys.\  B {\bf 629}, 3 (2002)
  [arXiv:hep-th/0112251].
  %%CITATION = NUPHA,B629,3;%%
}

%\SohniusPK
\lref\SohniusPK{
  M.~F.~Sohnius,
  ``The Multiplet Of Currents For N=2 Extended Supersymmetry,''
  Phys.\ Lett.\  B {\bf 81}, 8 (1979).
  %%CITATION = PHLTA,B81,8;%%
}

%\WittenZE
\lref\WittenZE{
  E.~Witten,
  ``Topological Quantum Field Theory,''
  Commun.\ Math.\ Phys.\  {\bf 117}, 353 (1988).
  %%CITATION = CMPHA,117,353;%%
}

%\StromingerPD
\lref\StromingerPD{
  A.~Strominger,
  ``SPECIAL GEOMETRY,''
  Commun.\ Math.\ Phys.\  {\bf 133}, 163 (1990).
  %%CITATION = CMPHA,133,163;%%
}

%\PeriwalMX
\lref\PeriwalMX{
  V.~Periwal and A.~Strominger,
  ``KAHLER GEOMETRY OF THE SPACE OF N=2 SUPERCONFORMAL FIELD THEORIES,''
  Phys.\ Lett.\  B {\bf 235}, 261 (1990).
  %%CITATION = PHLTA,B235,261;%%
}

\lref\KS{M.~Kontsevich and Y.~Soibelman,
``Stability Structures, Motivic Donaldson-Thomas Invariants and Cluster Transformations,"
[arXiv:0811.2435].}

%\DolanUT
\lref\DolanUT{
  F.~A.~Dolan and H.~Osborn,
  ``Conformal four point functions and the operator product expansion,''
  Nucl.\ Phys.\  B {\bf 599}, 459 (2001)
  [arXiv:hep-th/0011040].
  %%CITATION = NUPHA,B599,459;%%
}

\lref\wipk{Work in progress.}

\lref\BasuNT{
  A.~Basu, M.~B.~Green and S.~Sethi,
  ``Some systematics of the coupling constant dependence of N = 4
  Yang-Mills,''
  JHEP {\bf 0409}, 045 (2004)
  [arXiv:hep-th/0406231].
  %%CITATION = JHEPA,0409,045;%%
}

%\deBoerSS
\lref\deBoerSS{
  J.~de Boer, J.~Manschot, K.~Papadodimas and E.~Verlinde,
  ``The chiral ring of AdS3/CFT2 and the attractor mechanism,''
  JHEP {\bf 0903}, 030 (2009)
  [arXiv:0809.0507 [hep-th]].
  %%CITATION = JHEPA,0903,030;%%
}

%\SeibergPF
\lref\SeibergPF{
  N.~Seiberg,
  ``Observations On The Moduli Space Of Superconformal Field Theories,''
  Nucl.\ Phys.\  B {\bf 303}, 286 (1988).
  %%CITATION = NUPHA,B303,286;%%
}

%\KutasovXB
\lref\KutasovXB{
  D.~Kutasov,
  ``GEOMETRY ON THE SPACE OF CONFORMAL FIELD THEORIES AND CONTACT TERMS,''
  Phys.\ Lett.\  B {\bf 220}, 153 (1989).
  %%CITATION = PHLTA,B220,153;%%
}

%\MinwallaKA
\lref\MinwallaKA{
  S.~Minwalla,
  ``Restrictions imposed by superconformal invariance on quantum field
  theories,''
  Adv.\ Theor.\ Math.\ Phys.\  {\bf 2}, 781 (1998)
  [arXiv:hep-th/9712074].
  %%CITATION = 00203,2,781;%%
}

%\DolanZH
\lref\DolanZH{
  F.~A.~Dolan and H.~Osborn,
  ``On short and semi-short representations for four dimensional superconformal
  symmetry,''
  Annals Phys.\  {\bf 307}, 41 (2003)
  [arXiv:hep-th/0209056].
  %%CITATION = APNYA,307,41;%%
}
%\RanganathanNB
\lref\RanganathanNB{
  K.~Ranganathan,
  ``Nearby CFTs in the operator formalism: The Role of a connection,''
  Nucl.\ Phys.\  B {\bf 408}, 180 (1993)
  [arXiv:hep-th/9210090].
  %%CITATION = NUPHA,B408,180;%%
}

%\PolandWG
\lref\PolandWG{
  D.~Poland and D.~Simmons-Duffin,
  ``Bounds on 4D Conformal and Superconformal Field Theories,''
  arXiv:1009.2087 [hep-th].
  %%CITATION = ARXIV:1009.2087;%%
}

%\DolanTT
\lref\DolanTT{
  F.~A.~Dolan and H.~Osborn,
  ``Superconformal symmetry, correlation functions and the operator product
  expansion,''
  Nucl.\ Phys.\  B {\bf 629}, 3 (2002)
  [arXiv:hep-th/0112251].
  %%CITATION = NUPHA,B629,3;%%
}

%\RanganathanVJ
\lref\RanganathanVJ{
  K.~Ranganathan, H.~Sonoda and B.~Zwiebach,
  ``Connections on the state space over conformal field theories,''
  Nucl.\ Phys.\  B {\bf 414}, 405 (1994)
  [arXiv:hep-th/9304053].
  %%CITATION = NUPHA,B414,405;%%
}

%\KinneyEJ
\lref\KinneyEJ{
  J.~Kinney, J.~M.~Maldacena, S.~Minwalla and S.~Raju,
  ``An index for 4 dimensional super conformal theories,''
  Commun.\ Math.\ Phys.\  {\bf 275}, 209 (2007)
  [arXiv:hep-th/0510251].
  %%CITATION = CMPHA,275,209;%%
}

%\BershadskyCX
\lref\BershadskyCX{
  M.~Bershadsky, S.~Cecotti, H.~Ooguri and C.~Vafa,
  ``Kodaira-Spencer theory of gravity and exact results for quantum string
  amplitudes,''
  Commun.\ Math.\ Phys.\  {\bf 165}, 311 (1994)
  [arXiv:hep-th/9309140].
  %%CITATION = CMPHA,165,311;%%
}

%\deBoerQE
\lref\deBoerQE{ 
  J.~de Boer, K.~Papadodimas and E.~Verlinde,
  ``Black Hole Berry Phase,''
  arXiv:0809.5062 [hep-th].
  %%CITATION = ARXIV:0809.5062;%%
}

%\PedderWP
\lref\PedderWP{
  C.~Pedder, J.~Sonner and D.~Tong,
  %``The Geometric Phase and Gravitational Precession of D-Branes,''
  Phys.\ Rev.\  D {\bf 76}, 126014 (2007)
  [arXiv:0709.2136 [hep-th]].
  %%CITATION = PHRVA,D76,126014;%%
}

%\PedderFF
\lref\PedderFF{
  C.~Pedder, J.~Sonner and D.~Tong,
  %``The Geometric Phase in Supersymmetric Quantum Mechanics,''
  Phys.\ Rev.\  D {\bf 77}, 025009 (2008)
  [arXiv:0709.0731 [hep-th]].
  %%CITATION = PHRVA,D77,025009;%%
}
%\PedderJE
\lref\PedderJE{
  C.~Pedder, J.~Sonner and D.~Tong,
  %``The Berry Phase of D0-Branes,''
  JHEP {\bf 0803}, 065 (2008)
  [arXiv:0801.1813 [hep-th]].
  %%CITATION = JHEPA,0803,065;%%
}

%\SonnerFI
\lref\SonnerFI{
  J.~Sonner and D.~Tong,
  %``Berry Phase and Supersymmetry,''
  JHEP {\bf 0901}, 063 (2009)
  [arXiv:0810.1280 [hep-th]].
  %%CITATION = JHEPA,0901,063;%%
}

%\ShapereZF
\lref\ShapereZF{
  A.~D.~Shapere and Y.~Tachikawa,
  ``Central charges of N=2 superconformal field theories in four dimensions,''
  JHEP {\bf 0809}, 109 (2008)
  [arXiv:0804.1957 [hep-th]].
  %%CITATION = JHEPA,0809,109;%%
}

%\LercheUY
\lref\LercheUY{
  W.~Lerche, C.~Vafa and N.~P.~Warner,
  ``Chiral Rings in N=2 Superconformal Theories,''
  Nucl.\ Phys.\  B {\bf 324}, 427 (1989).
  %%CITATION = NUPHA,B324,427;%%
}

%\SonodaDH
\lref\SonodaDH{
  H.~Sonoda,
  ``Connection on the theory space,''
  arXiv:hep-th/9306119.
  %%CITATION = HEP-TH/9306119;%%
}

%\ArgyresJJ
\lref\ArgyresJJ{
  P.~C.~Argyres and M.~R.~Douglas,
  ``New phenomena in SU(3) supersymmetric gauge theory,''
  Nucl.\ Phys.\  B {\bf 448}, 93 (1995)
  [arXiv:hep-th/9505062].
  %%CITATION = NUPHA,B448,93;%%
}

%\ArgyresXN
\lref\ArgyresXN{
  P.~C.~Argyres, M.~Ronen Plesser, N.~Seiberg and E.~Witten,
  ``New N=2 Superconformal Field Theories in Four Dimensions,''
  Nucl.\ Phys.\  B {\bf 461}, 71 (1996)
  [arXiv:hep-th/9511154].
  %%CITATION = NUPHA,B461,71;%%
}

%\SonodaMV
\lref\SonodaMV{
  H.~Sonoda,
  ``Composite operators in QCD,''
  Nucl.\ Phys.\  B {\bf 383}, 173 (1992)
  [arXiv:hep-th/9205085].
  %%CITATION = NUPHA,B383,173;%%
}

%\ArgyresCN
\lref\ArgyresCN{
  P.~C.~Argyres and N.~Seiberg,
  ``S-duality in N=2 supersymmetric gauge theories,''
  JHEP {\bf 0712}, 088 (2007)
  [arXiv:0711.0054 [hep-th]].
  %%CITATION = JHEPA,0712,088;%%
}

%\GaiottoGZ
\lref\GaiottoGZ{
  D.~Gaiotto and J.~Maldacena,
  ``The gravity duals of N=2 superconformal field theories,''
  arXiv:0904.4466 [hep-th].
  %%CITATION = ARXIV:0904.4466;%%
}

%\TachikawaTQ
\lref\TachikawaTQ{
  Y.~Tachikawa,
  ``Five-dimensional supergravity dual of a-maximization,''
  Nucl.\ Phys.\  B {\bf 733}, 188 (2006)
  [arXiv:hep-th/0507057].
  %%CITATION = NUPHA,B733,188;%%
}
%\AldayAQ
\lref\AldayAQ{
  L.~F.~Alday, D.~Gaiotto and Y.~Tachikawa,
  ``Liouville Correlation Functions from Four-dimensional Gauge Theories,''
  arXiv:0906.3219 [hep-th].
  %%CITATION = ARXIV:0906.3219;%%
}

%\AldayFS
\lref\AldayFS{
  L.~F.~Alday, D.~Gaiotto, S.~Gukov, Y.~Tachikawa and H.~Verlinde,
  ``Loop and surface operators in N=2 gauge theory and Liouville modular
  geometry,''
  arXiv:0909.0945 [hep-th].
  %%CITATION = ARXIV:0909.0945;%%
}

%\IntriligatorJJ
\lref\IntriligatorJJ{
  K.~A.~Intriligator and B.~Wecht,
  ``The exact superconformal R-symmetry maximizes a,''
  Nucl.\ Phys.\  B {\bf 667}, 183 (2003)
  [arXiv:hep-th/0304128].
  %%CITATION = NUPHA,B667,183;%%
}
%\CecottiKZ
\lref\CecottiKZ{
  S.~Cecotti,
  ``N=2 Landau-Ginzburg versus Calabi-Yau sigma models: Nonperturbative
  aspects,''
  Int.\ J.\ Mod.\ Phys.\  A {\bf 6}, 1749 (1991).
  %%CITATION = IMPAE,A6,1749;%%
}

%\CecottiUF
\lref\CecottiUF{
  S.~Cecotti and C.~Vafa,
  ``BPS Wall Crossing and Topological Strings,''
  arXiv:0910.2615 [hep-th].
  %%CITATION = ARXIV:0910.2615;%%
}

%\DijkgraafDJ
\lref\DijkgraafDJ{
  R.~Dijkgraaf, H.~L.~Verlinde and E.~P.~Verlinde,
  ``Topological Strings In D $<$ 1,''
  Nucl.\ Phys.\  B {\bf 352}, 59 (1991).
  %%CITATION = NUPHA,B352,59;%%
}

%\DijkgraafQW
\lref\DijkgraafQW{
  R.~Dijkgraaf, H.~L.~Verlinde and E.~P.~Verlinde,
  ``Notes on topological string theory and 2-D quantum gravity,''
  %%CITATION = C90-05-27;%%
}

%\MarshakovAE
\lref\MarshakovAE{
  A.~Marshakov, A.~Mironov and A.~Morozov,
  ``WDVV-like equations in N = 2 SUSY Yang-Mills theory,''
  Phys.\ Lett.\  B {\bf 389}, 43 (1996)
  [arXiv:hep-th/9607109].
  %%CITATION = PHLTA,B389,43;%%
}

%\MarshakovNY
\lref\MarshakovNY{
  A.~Marshakov, A.~Mironov and A.~Morozov,
  ``WDVV equations from algebra of forms,''
  Mod.\ Phys.\ Lett.\  A {\bf 12}, 773 (1997)
  [arXiv:hep-th/9701014].
  %%CITATION = MPLAE,A12,773;%%
}

%\MarshakovCR
\lref\MarshakovCR{
  A.~Marshakov, A.~Mironov and A.~Morozov,
  ``More evidence for the WDVV equations in N = 2 SUSY Yang-Mills theories,''
  Int.\ J.\ Mod.\ Phys.\  A {\bf 15}, 1157 (2000)
  [arXiv:hep-th/9701123].
  %%CITATION = IMPAE,A15,1157;%%
}

%\WittenIG
\lref\WittenIG{
  E.~Witten,
  ``ON THE STRUCTURE OF THE TOPOLOGICAL PHASE OF TWO-DIMENSIONAL GRAVITY,''
  Nucl.\ Phys.\  B {\bf 340}, 281 (1990).
  %%CITATION = NUPHA,B340,281;%%
}

%\PestunRZ
\lref\PestunRZ{
  V.~Pestun,
  ``Localization of gauge theory on a four-sphere and supersymmetric Wilson
  loops,''
  arXiv:0712.2824 [hep-th].
  %%CITATION = ARXIV:0712.2824;%%
}

\lref\KolZT{
  B.~Kol,
  ``On conformal deformations,''
  JHEP {\bf 0209}, 046 (2002)
  [arXiv:hep-th/0205141].
  %%CITATION = JHEPA,0209,046;%%
}

\lref\Helgason{
  S. Helgason,
 ``Differential Geometry and Symmetric Spaces'',
  Academic Press, 196}

\lref\BonelliQH{
  G.~Bonelli and M.~Matone,
  ``Nonperturbative Relations in N=2 Susy Yang-Mills and WDVV Equation,''
  Phys.\ Rev.\ Lett.\  {\bf 77}, 4712 (1996)
  [arXiv:hep-th/9605090].
  %%CITATION = PRLTA,77,4712;%%
}
\lref\GaiottoCD{
  D.~Gaiotto, G.~W.~Moore and A.~Neitzke,
  ``Four-dimensional wall-crossing via three-dimensional field theory,''
  arXiv:0807.4723 [hep-th].
  %%CITATION = ARXIV:0807.4723;%%
}

\Title{\vbox{\baselineskip12pt}}
{\vbox{\centerline {  Topological Anti - Topological Fusion}
\smallskip
  \centerline{in Four-Dimensional Superconformal Field Theories}}}
\centerline{Kyriakos Papadodimas}

\bigskip
\centerline{ Institute for Theoretical Physics}
\centerline{University of Amsterdam}
\centerline{Valckenierstraat 65, 1018 XE}
\centerline{Amsterdam, The Netherlands}
\centerline{\it k.papadodimas@uva.nl}
\bigskip
\medskip

\bigskip

\noindent We present some new exact results for general four-dimensional 
superconformal field theories.  We derive differential equations
governing the coupling constant dependence of chiral primary
correlators. For ${\cal N}=2$ theories we show that the Zamolodchikov
metric on the moduli space and the operator mixing of chiral primaries
are quasi-topological quantities and constrained by holomorphy.  The
equations that we find are the four-dimensional analogue of the $tt^*$
equations in two-dimensions, discovered by the method of ``topological
anti-topological fusion'' by Cecotti and Vafa.  Our analysis relies on
conformal perturbation theory and the superconformal Ward identities
and does not use a topological twist.

\Date{}

\newsec{Introduction}

Supersymmetric field theories in four dimensions are interesting for
their possible phenomenological applications and as toy models for the
analysis of non-perturbative phenomena in quantum field theory.
Supersymmetry allows us to go beyond perturbation theory and to
derive exact results, which proved important for the analysis of
strongly coupled gauge theories and the discovery of dualities.

In this paper we will derive some new exact results for general
four-dimensional superconformal field theories. We will mainly focus
on theories with ${\cal N}=2$ supersymmetry, but some of our results
are also true for ${\cal N}=1$ theories. A basic example of ${\cal
N}=2$ SCFTs in four dimensions is the ${\cal N}=2$ $SU(N)$ gauge
theory with $N_f=2N$ massless quarks in the fundamental
representation \SeibergAJ,\ArgyresWT. Other ${\cal N}=2$ theories in
four dimensions
include \ArgyresJJ,\ArgyresXN,\EguchiVU,\EguchiDS,\MinahanFG,\MinahanCJ,\AharonyXZ,\ArgyresPP,\ArgyresWX,\ArgyresTQ.
More recently a larger class of related theories has been studied by
Gaiotto \GaiottoWE. They arise by compactifying the six-dimensional
superconformal field theory living on $N$ M5 branes on a punctured
Riemann surface. Our discussion will be general and not based on any
specific theory.

Many superconformal theories in four dimensions are not isolated, but
come in continuous families parametrized by coupling constants, which
can be freely adjusted without breaking conformal invariance.  The set
of possible values for these coupling constants often has the
structure of a smooth manifold that we call the moduli space ${\cal
M}$ of the conformal field theory. This space plays the role of a
``parameter space'' and has to be distinguished from the moduli space
of vacua, like the Coulomb or Higgs branch, of any specific conformal
field theory\foot{For example in the case of the ${\cal N}=4$ super
Yang-Mills the moduli space is the upper half plane parametrized by
$\tau = {\theta \over 2\pi} + i {4 \pi \over g_{_{YM}}^2}$, modded out
by the action of the $SL(2,Z)$ duality group.}.  Motion along ${\cal
M}$ is generated by perturbing the conformal field theory by marginal
operators. In general there may be special points on the moduli space
where the conformal field theory admits a weakly coupled Lagrangian
description, but at a generic point of ${\cal M}$ the theory is
strongly coupled.

Our main goal is to develop a method that will allow us to probe the
interior of the moduli space non-perturbatively. For this we will
derive a set of classical differential equations for correlators of
BPS operators, when considered as functions of the coupling constants.
The value of such correlators in the weak-coupling regions of the
moduli space provide ``boundary conditions'' for these differential
equations, whose solutions can then give us the correlators at all
values of the coupling.  This is potentially useful because it may
allow us to relate weak and strong coupling results and to probe the
interior of the moduli space, where no weakly coupled description is
available. The BPS operators that we will study are the chiral
primaries with their ``chiral ring'' multiplication. Additionally we
want to understand how superconformal invariance constrains the
Zamolodchikov metric on the moduli space.

In two-dimensional ${\cal N}=(2,2)$ theories many exact results are
known about the correlation functions of BPS operators. By considering
the topologically twisted theory it was discovered that such
correlators obey differential equations with respect to the coupling
constants, called the Witten-Dijkgraaf-Verlinde-Verlinde (WDVV)
equations \DijkgraafDJ,\DijkgraafQW,\WittenIG. These equations govern
the dependence of correlators with respect to holomorphic deformations
of the couplings. While useful in topological field theories, it is
not easy to directly apply these equations to the physical quantum
field theory since for a deformation to be consistent with unitarity
it must be a real linear combination of holomorphic and
antiholomorphic deformations.

In \CecottiME\ a beautiful method, called ``topological
anti-topological fusion'', was developed by Cecotti and Vafa, which
allows one to consider simultaneous holomorphic and antiholomorphic
deformations of correlators in two-dimensional ${\cal N}=(2,2)$
theories. Such correlators can be computed by gluing
together a hemisphere on which the the theory is topologically twisted
and one where the conjugate twisting is performed. The resulting
amplitude turns out to be quasi-topological and constrained to
satisfy exact differential equations called the $tt^*$ equations. We
would like to emphasize that while the derivation was based on the
topologically twisted theory, the final equations are valid even in
the physical, untwisted quantum field theory.
 
How much of this structure survives in higher dimensional supersymmetric field theories?
At first thought it appears to be difficult to generalize the 
arguments of \CecottiME\ to four-dimensional superconformal field theories. Some
important ingredients of the two-dimensional story, such as the
correspondence between Ramond ground states and chiral primaries via
spectral flow are missing in four-dimensions. However in \deBoerSS\
it was shown that in the case of two-dimensional ${\cal N}=(2,2)$
superconformal field theories, the $tt^*$ equations can also be derived
from the point of view of standard conformal perturbation theory in
the NS sector, without relying on the topological twisting. We will
show that with small modifications the same arguments can be applied
to four-dimensional ${\cal N}=2$ superconformal field theories.

To summarize our results we find that four-dimensional ${\cal N}=2$
superconformal field theories have rich mathematical structure,
comparable to that of ${\cal N}=(2,2)$ theories in two dimensions, as
far their chiral ring sector is concerned. We find that the
Zamolodchikov metric on the moduli space ${\cal M}$ of a four-dimensional ${\cal N}=2$ theory satisfies constraints similar to those
of ``special geometry''.

For ${\cal N}=1$ and ${\cal N}=2$ theories we show that the 3-point
functions $C_{ij}^k$ of chiral primary operators vary holomorphically
on ${\cal M}$ and satisfy the WDVV integrability equations. Similarly
more general ``extremal correlators'' of chiral primaries vary holomorphically.
Finally for ${\cal N}=2$ theories we find that the operator mixing for
chiral primaries is characterized by the fact that such operators are
sections of holomorphic vector bundles over ${\cal M}$ whose curvature
is computed by the $tt^*$ equations. As we will explain in the main
text, these can be considered as a partial nonlinear differential
equation relating the 2- and 3-point functions of chiral primaries
over the moduli space. While the 2-point functions $g_{k\overline{l}}$
are not holomorphic, they are computed as solutions of these
differential equations, whose coefficients are the holomorphic
functions $C_{ij}^k$. 

In this paper we will focus on the basic derivation of the
four-dimensional $tt^*$ equations from conformal perturbation
theory. We hope to report on possible applications and on the relation
to the topologically twisted theory in the future \wipk.

The plan of the paper is as follows: in section 2 we briefly summarize
some basic properties of ${\cal N}=2$ superconformal field theories in
four dimensions. In section 3 we discuss aspects of conformal
perturbation theory and operator mixing. In section 4 we present
our main results for ${\cal N}=2$ theories. In section 5 we describe
the proof of the $tt^*$ equations.  In section 6 we analyze the
constraints for the Zamolodchikov metric on the moduli space. In
section 7 we consider our results for ${\cal N}=1$ superconformal
field theories. In section 8 we make some observations about ${\cal
N}=4$ theories and in section 9 we make some remarks about the space
of vacua of such theories. In in section 10 we have some general
comments and discuss possible applications and extensions.

\newsec{${\cal N}=2$ superconformal field theories in four dimensions}

In this section we review some basic facts about four-dimensional
${\cal N}=2$ superconformal field theories. Some useful references
for the algebra and its representations are \MinwallaKA,\DolanZH.

\subsec{The algebra and the chiral ring}
 The ${\cal N}=2$ superconformal algebra consists of the standard
conformal generators $P_\mu,K_\mu,M_{\mu\nu},D$, the 8 supercharges
$Q_a^i,\overline{Q}_{i,\dot{a}}$ and their superconformal partners
$S^a_i, \overline{S}^{i,\dot{a}}$. Additionally it contains the
generators of the $SU(2)_R\times U(1)_R$ R-symmetry algebra.  Here
$a,\dot{a}$ are Lorentz spinor indices and $i$ is an $SU(2)_R$ index
in the $I={1\over 2}$ representation. We work in conventions where the
left chiral supercharges $Q_a^i$ have $U(1)_R$ charge equal to
$-1$. Superconformal primary operators\foot{We call conformal primary
operators those which are annihilated by the $K_\mu$'s. Superconformal
primary operators are those which annihilated by all $S$ and
$\overline{S}$'s, which implies that they are also annihilated by the
$K_\mu$'s.} are labeled by their conformal dimension $\Delta$, the
Lorentz spin $(j,\overline{j})$, the $SU(2)_R$ ``isospin'' $I$ and the
$U(1)_R$ charge $R$.

We will be interested in chiral primary operators $\bphi$ which
satisfy the following conditions
$$  
[\overline{Q}^i_{\dot{a}},\bphi] = 0 ,\qquad \Leftrightarrow \qquad \overline{j}=0 \,, \,I=0,\, \Delta = {R\over 2},
$$
The antichiral operators $\overline{\bphi}$ have
$j=0,\, I=0,\, \Delta=-{R\over 2}$ and are annihilated by the supercharges
of left chirality $Q_a^i$. Notice that we defined the ``chiral
primaries'' as operators which are
annihilated by all supercharges of one chirality, or in other words,
which are chiral with respect to both ${\cal N}=1$
sub-algebras of the ${\cal N}=2$ theory. This definition implies $I=0$.

The 2-point function of chiral primaries defines the Zamolodchikov metric for
these operators\foot{When we write operators at infinity, such as $\overline{\bphi}_j(\infty)$, what we really mean is explained in equation (2.2).}
\vskip-20pt
$$
 g_{i\overline{j}}\equiv \langle \overline{\bphi}_j(\infty) \bphi_i(0) \rangle
$$
Using the $U(1)_R$ conservation it is easy to show that the operator
product expansion of chiral primaries is non-singular
\vskip-20pt
$$
\bphi_i(x) \bphi_j(0) = C^i_{jk} \bphi_k(0) + ...
$$
and that the operator $\bphi_k(0)$ is also chiral primary of charge
$R_k=R_i+R_j$. In this sense the chiral primary operators form a ring
under OPE multiplication, called the ``chiral ring'' \LercheUY. The
constants $C_{ij}^k$ are the structure constants of the ring. The
chiral ring coefficients are related to the 2- and 3-point functions
of chiral primaries by\foot{Any three points can be mapped to
$\{0,1,\infty\}$ by a conformal transformation, so the 3-point
function is fixed by conformal invariance up to an overall constant.}
$$
    C_{ij\overline{k}}\equiv\langle\overline{\bphi}_k(\infty) \bphi_i(1)\bphi_j(0)\rangle 
$$
$$
\qquad C_{ij\overline{k}} = C_{ij}^l g_{l\overline{k}}
$$

\subsec{Superconformal Ward identities}

In the rest of the paper we will often use superconformal Ward
identities so we briefly review them here. A basic Ward identity is
that for a set of local bosonic operators $\gop_i$ and a supercharge
${\bf Q}$ we have
\eqn\basicward{
\sum_{k=1}^n\langle\gop _1 (x_1)...[{\bf Q},\gop_k](x_k)...\gop_n(x_n) 
\rangle= 0}
which expresses the ``supercharge conservation''. For simplicity in
this section we will assume that all local operators $\varphi_i$ are
bosonic\foot{For fermionic operators there may be additional minus
signs in the Ward identities.}.

We will also need a somewhat less familiar identity. First we remind
that if $\gop$ is a scalar\foot{For operators with spin we have to
multiply with the inverse 2-point function.} conformal primary
operator of dimension $\Delta$ then we define
\eqn\infop{
\langle\gop(\infty)\gop_1(x_1)...\gop_n(x_n) \rangle \equiv \lim_{x\rightarrow
\infty} |x|^{2\Delta}\langle \gop(x) \gop_1(x_1)...\gop_n(x_n) \rangle}
Now let us consider the correlator of a superconformal primary
operator $\gop$ with a number of operators $\gop_i$ (not necessarily
primary). According to \basicward\ we have
$$
\langle [{\bf Q},\gop](x) \gop_1(x_1)...\gop_n(x_n)\rangle +
\sum_{k=1}^n\langle \gop(x)\gop_1 (x_1)...[{\bf Q},\gop_k](x_k)...\gop_n(x_n) 
\rangle= 0
$$
We multiply this relation by $|x|^{2\Delta}$ where $\Delta$ is the
dimension of $\varphi$ and take the limit $x\rightarrow \infty$. The
term $|x|^{2\Delta}\langle [{\bf
Q},\gop](x) \gop_1(x_1)...\gop_n(x_n)\rangle$ goes to zero in that
limit, because $[{\bf Q},\gop](x)$ is a conformal primary\foot{From
the ${\cal N}=2$ algebra we have $[K_\mu,{\bf Q}] \sim
{\bf \overline{S}}$, so since $\gop$ is a superconformal primary we
have $[K_\mu,\gop]=[{\bf \overline{S}},\gop]=0$ and thus $[K_\mu,[{\bf
Q},\gop]]=0$.} of dimension $\Delta+{1\over 2}$ so this term falls off
at least as fast as ${1\over |x|}$ while the other terms have a finite
limit according to \infop\ and we have
\eqn\basicwardb{
\sum_{k=1}^n\langle \gop(\infty)\gop_1 (x_1)...[{\bf Q},\gop_k](x_k)...\gop_n(x_n) 
\rangle= 0}
where in this sum there is no term where the supercharge ${\bf Q}$
acts on $\phi(\infty)$.  So when applying the superconformal Ward
identity \basicward\ if a superconformal primary is ``hidden'' at
infinity then ${\bf Q}$ does not act on it.

Of course there is nothing special about the point at infinity, since
a conformal transformation can map it to any other point.  The
identity \basicwardb\ is a special case of a more general Ward
identity, which holds even if all insertions in the correlator are at
finite locations: for a given superconformal primary $\varphi$ in the
correlator, it is always possible to find a linear combination of Ward
identities (for $Q$'s and $\overline{S}$'s), in which there are no
terms involving operators of the form $[Q,\varphi]$. This is achieved
by applying the Ward identity for the supercurrent multiplied by a
conformal Killing spinor which vanishes at the location of
$\varphi$. Above we presented the special case $x\rightarrow \infty$ because
then the argument is quicker. The general case is explained in
appendix A.

\newsec{Marginal Deformations and Operator Mixing}

In ${\cal N}=2$ superconformal field theories exactly marginal
operators preserving supersymmetry must be descendants of scalar
chiral primaries of dimension $\Delta=2$, $U(1)_R$ charge $R= 4$ and
their conjugates. In the rest of the paper we denote chiral primaries
of this R-charge by capital letters $\Phi_i$, while for chiral
primaries of general charge we use the notation $\phi_i$.

The moduli space has the structure of a complex K\"ahler manifold. The
marginal operators are divided into holomorphic and antiholomorphic
ones as\foot{More precisely we have $ {\cal
O}_i \propto \epsilon_{rm}\epsilon_{sn}\epsilon^{cd} \epsilon^{ab}\{Q^r_c,
[Q^s_d, \{Q^m_a,[Q^n_b,\lphi_i]\}] $ where the constant of
proportionality is a matter of conventions and can be adjusted to
simplify certain equations.}
\eqn\margoper{
{\cal O}_i = Q^4\cdot \lphi_i \qquad ,\qquad \overline{\cal O}_j
= \overline{Q}^4\cdot \overline{\lphi}_j} In superspace language the
two deformations can be written as ${\cal N}=2$ F-terms of the form
$$
{\cal O}_i = \int d^4\theta \lphi_i\qquad ,\qquad \overline{\cal O}_j
= \int d^4 \overline{\theta} \,\,\overline{\lphi}_j
$$
The 2-point functions of chiral primaries and those of marginal
operators
\eqn\metriczam{
  \langle \lphi_i(x) \overline{\lphi}_j(y) \rangle =
  {g_{i\overline{j}} \over |x-y|^4},\qquad
\langle {\cal O}_i(x)\overline{\cal O}_j(y) \rangle =  {G_{i\overline{j}} \over |x-y|^8}}
can be related by using \basicward\ to move the $Q$'s from ${\cal
O}_i$ onto $\overline{\cal O}_j$ and then using the supersymmetry
algebra $\{Q_a^i,\overline{Q}_{j,\dot{b}}\} = 2 \delta^i_j
P_{a\dot{b}}$. With appropriate normalization of \margoper\ we have
\eqn\normalr{
\langle {\cal O}_i(x)\overline{\cal O}_j(y) \rangle  = 
\Boxy\Boxy\langle \lphi_i(x) \overline{\lphi}_j(y) \rangle}
where $\Boxy$ is the Laplacian operator with respect to $y$. This expression implies $G_{i\overline{j}} = 192\cdot g_{i\overline{j}}$. The
quantity $G_{i\overline{j}}$ is the Zamolodchikov metric on the moduli
space ${\cal M}$.

Now we want to analyze how the chiral ring varies as a function of the
coupling constants $\{\lambda^\mu\}$ parametrizing ${\cal M}$. At
first one may think that the coupling constant dependence is simply
captured by the fact that the 3-point functions of chiral primaries
become moduli-dependent functions $C_{ij}^k(\lambda)$. However the
situation is somewhat more complicated due to the operator mixing of
chiral primaries with the same quantum numbers under marginal
deformations.

To understand this better, let us first notice that at each point on
the moduli space we can choose the basis of chiral primaries in an
arbitrary way.  Under a coupling constant dependent change of basis
$\phi_i(\lambda) \rightarrow U_i^{i'}(\lambda) \phi_{i'}(\lambda)$ the
3-point functions transform as $C_{ij}^k(\lambda) \rightarrow
U_i^{i'}(\lambda)U_j^{j'}(\lambda) U_{k'}^{k}(\lambda)
C_{i'j'}^{k'}(\lambda)$.  From this ambiguity it is clear that
comparing the chiral ring i.e. the structure constants $C_{ij}^k$ or
the 3-point functions $C_{ij\overline{k}}$ at different points on
${\cal M}$ is not completely straightforward.

Is there a canonical way to choose a basis of operators over the
moduli space and to eliminate this ambiguity? As we now explain the
answer is negative. Operator mixing is an intrinsic property of the
conformal field theory, completely determined by its dynamics, and it
is impossible to gauge it away by an appropriate choice of basis.

One way to understand why the operator mixing is unavoidable is the
following: under an infinitesimal deformation by a marginal operator
${\cal O}$ the correlators of the conformal field theory change by\foot{We
included the factor of ${1\over (2\pi)^2}$ in the definition of the marginal 
deformations in order to simplify certain formulas.}
\eqn\divdef{
{1\over (2\pi)^2}\int d^4 x \langle {\cal O}(x) \varphi_1(z_1)...\varphi_n(z_n)\rangle }
In general these integrated correlators are divergent due to short
distance singularities in the OPE between ${\cal O}$ and the other
insertions.  This means that to define the deformed correlators we
have to find a way to renormalize the divergent integral \divdef.

The renormalization method that we use is quite
natural \RanganathanVJ: to first order in the perturbation we cut out
small balls of radius $\varepsilon$ around the insertions
$\varphi_i(x_i)$, and we compute the integrated correlator as a
function of the cutoff $\varepsilon$. Then we expand the integrated
correlator in powers of ${1 \over \varepsilon}$, throw away all the
divergent pieces and keep the finite result as $\varepsilon\rightarrow
0$. This defines ``renormalized'' deformed correlators to first order
in the deformation.  Going to second order we basically do the same,
but we have to be careful about the pieces that we already subtracted
at first order.  This approach was discussed
in \SonodaMV, \SonodaDH, \RanganathanNB\ and nicely developed for
deformations of two-dimensional conformal field theories
in \RanganathanVJ, to which we refer the reader for more details.

A careful analysis shows that if we apply this procedure twice
to compute the second order perturbation we find that in general the
following expression does not vanish
\eqn\curvatu{
{1\over (2\pi)^4}\int d^4 x\int d^4y \langle {\cal O_{[\mu}}(x) {\cal O_{\nu]}}(y)\varphi_1(z_1)...\varphi_n(z_n)\rangle\neq 0}
where the brackets denote antisymmetrization. The non-vanishing of this expression
indicates that the two deformations do not commute, i.e.  there is
curvature on the space of operators. In other words, the operator
mixing is inevitable because there is no way one can consistently
renormalize the deformed correlators and at the same time
avoid \curvatu\foot{Sometimes this discussion is presented in terms of
``contact terms'' in the OPE between ${\cal O}$ and the other
operators. The contact terms are tuned in such a way as to cancel the
divergent parts of the integrated correlator, but may also give finite
contributions which are interpreted as the connection on the space of
operators \SeibergPF,\KutasovXB.

In a sense the contact terms are not an intrinsic notion in a fixed
CFT, before we consider perturbing it. In a given CFT only correlators
of operators at distinct points are meaningful. The contact terms have
to be introduced when we want to consider integrated correlators
which, as we discussed, require the use of some renormalization
method.  The contact terms are then chosen in such a way that they
impose our preferred renormalization procedure in
all correlators in a consistent way, and allow us to do the integrals
such as \divdef\ including the coincident points.

On the other hand if we decide to explicitly use our renormalization
prescription \RanganathanNB\ whenever we deal with deformed
correlators, then we will never have to bring two operators at exactly
the same point and we do not need to worry about contact terms. Said
differently, the contact terms do not have to be included in the
renormalization method, but rather they are determined by it. In the
rest of the paper we will follow the prescription of \RanganathanVJ\
which was described above and will work in terms of regularized and
renormalized integrated correlators, so we will not have to talk about
contact terms anymore. We would like to thank K. Skenderis for
discussions about the contact terms.}.

Because of the operator mixing we should be thinking of chiral
primaries as operators taking values in vector bundles over the moduli
space of the theory. These bundles have nontrivial curvature. Let us
call ${\cal V}_R$ the bundle of the chiral primaries of $U(1)_R$
charge $R$. Then the chiral ring coefficients describe the OPE
multiplication between different bundles
$$
  C^k_{ij} : {\cal V}_R \otimes {\cal V}_{R'} \rightarrow {\cal V}_{R+R'}
$$
Each of the ${\cal V}_R$ bundles has nontrivial connection which
encodes the operator mixing. For a set of chiral primaries of the same
charge $\phi_i$ which can mix under motion on ${\cal M}$ we describe
the mixing by the connection $A_\mu$ and we have the covariant
derivative and its curvature
$$
  \nabla_\mu  = \partial_\mu  + A_\mu ,\qquad F_{\mu\nu} \equiv [\nabla_\mu,\nabla_\nu]
$$
%$$
%F_{\mu\nu} \equiv [\nabla_\mu,\nabla_\nu]  %\partial_\mu A_\nu - \partial_\nu A_\mu + [A_\mu,A_\nu]
%$$
The connection is determined by the dynamics of the conformal field
theory by the following condition
\eqn\defconn{
\nabla_\mu \langle \varphi_1(z_1)...\varphi_n(z_n)\rangle = {1\over (2\pi)^2}\left[
  \int d^4 x \,\langle {\cal
  O}_\mu(x) \varphi_1(z_1)...\varphi_n(z_n)\rangle \right]_{renormalized}}
  where $\varphi_i$ are general local operators (not necessarily
  chiral primary) and the renormalization prescription is the one we
  mentioned earlier. From now on, whenever we write integrated
  correlators we will always refer to their ``renormalized'' values,
  so we will drop the ``renormalized'' subscript from the integrals.
\fig{Computation of operator mixing from a 4-point function.}
{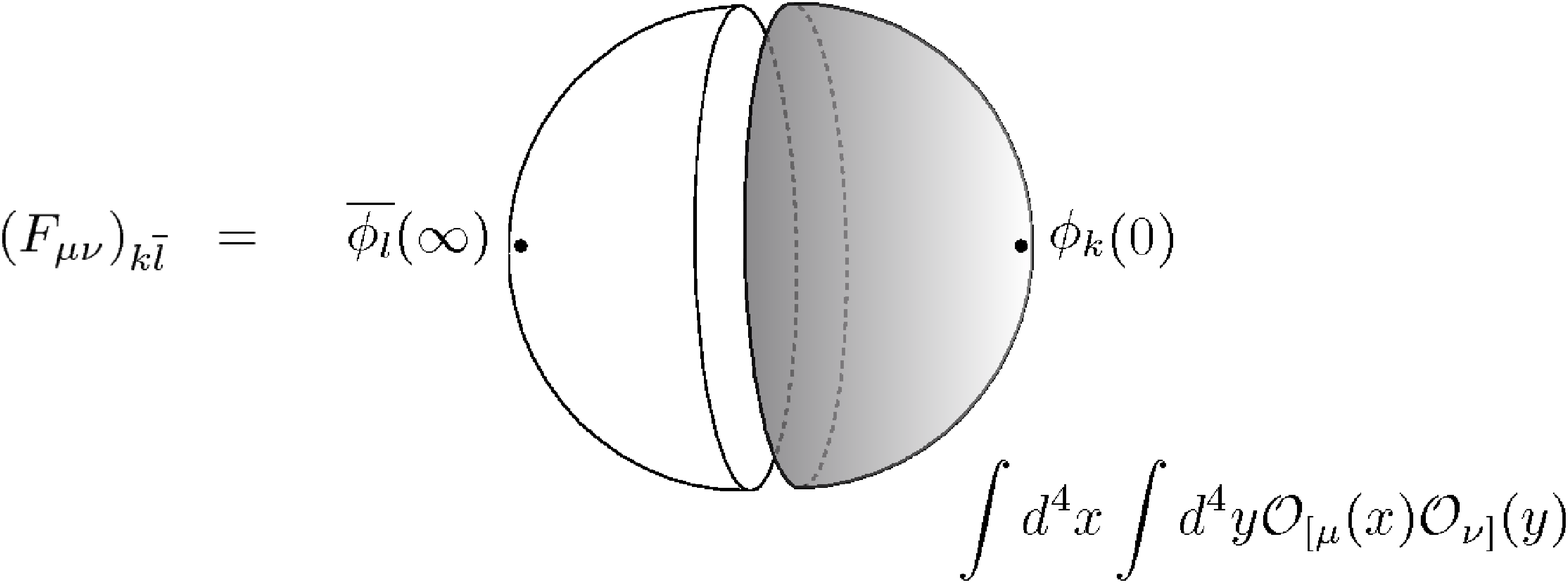}{5.truein}
\figlabel{\curvature}
Starting from \defconn\ it is straightforward to find an expression
for the curvature of the bundle of chiral primary operators. We
consider deformations by two marginal operators ${\cal O}_\mu,{\cal
O}_\nu$ and compute the infinitesimal variation of an operator
$\phi_k$ under the deformation by the antisymmetrized combination
${\cal O}_{[\mu}(x) {\cal O}_{\nu]}(y)$. Then we project this
variation on the space of operators with which $\phi_k$ mixes.  In
terms of correlators this can be computed by taking the 4-sphere ${\bf
S}^4$, dividing it into two hemispheres, inserting $\phi_k$ at the
center of one hemisphere and the conjugate operator
$\overline{\phi}_l$ on the other hemisphere and then integrating the
combination $\int d^4x \int d^4y {\cal O}_{[\mu}(x) {\cal
O}_{\nu]}(y)$ on the hemisphere where $\phi_k$ was inserted, as
depicted in figure \curvature. Mapping the ${\bf S}^4$ onto ${\bf
R}^4$ we find that the curvature can also be written as
\eqn\maincurv{
(F_{\mu\nu})_{k\overline{l}} = {1\over (2\pi)^4}\int_{|x|\leq 1} d^4x \int_{|y|\leq 1}
d^4y \langle \overline{\phi}_l(\infty) {\cal O}_{[\mu}(x) {\cal
O}_{\nu]}(y) \phi_k(0)\rangle} where again the double integral is
supposed to be renormalized by the prescription that we mentioned
earlier, see also \RanganathanVJ, \deBoerSS\ for more details. From
the curvature it is possible to reconstruct the connection $A_{\mu
i}^j$, at least locally. Once this connection has been computed, then
it is meaningful to ask what is the dependence of the chiral ring
coefficients on the coupling constants.
 
To give a simple example why all of the previous analysis was
necessary, let us consider a theory where we believe that the 3-point
functions of chiral primaries are ``independent of the coupling
constant''. According to what we discussed this statement should be
interpreted as $\nabla_\mu C_{ij}^k = 0$ and not $\partial_\mu
C_{ij}^k=0$ which is a non-covariant, basis-dependent equation. What
this means in practice is that in a theory where $F_{\mu\nu}\neq 0$,
it is impossible to choose a basis such that $\partial_\mu
C_{ij}^k=0$ everywhere on the moduli space, while it is
still possible to have $\nabla_\mu C_{ij}^k=0$. This is the right way
to express the notion that the 3-point functions are ``independent of
the coupling constant''.

Let us make a few more comments. In all of the above we have assumed
that the conformal dimensions of the operators involved in the
correlators do not change as we vary the coupling constants. This
assumption was motivated because we are interested in correlation
functions of chiral primary operators, whose conformal dimension
cannot change as long as they remain ``chiral
primaries''\foot{Consider an operator $\phi$ which is chiral primary
everywhere on the moduli space, so it satisfies $\Delta= {R\over 2}$
everywhere. Can its R-charge be a function of the coupling constants?
Since it is an abelian charge one may wonder whether it can change
continuously and simultaneously with $\Delta$, preserving the
condition $\Delta={R \over 2}$. However it is easy to use a
superconformal Ward identity to show that the 3-point function
$\langle{\cal O} \phi \overline{\phi}\rangle$ vanishes, which means
that the conformal dimension (and hence R-charge) of a chiral primary
cannot change continuously on the moduli space. Notice that the
discussion in this paragraph is not about whether chiral primaries can
combine into long multiplets and lift from the unitarity bound $\Delta
= {R\over 2}$, which is not excluded by this argument.  We have
ignored this effect because it can only happen at special points of
the moduli space. We would like to thank J. de Boer for discussions on
these issues. }. If we consider more general operators whose conformal
dimensions vary continuously, then the deformed correlators can have
logarithmic terms and we have to be more careful in defining the
connection.

Finally, the connection defined above is compatible with the
Zamolodchikov metric defined by the 2-point functions of chiral
primaries
\eqn\compat{
\nabla_\mu g_{k\overline{l}} \equiv 
\nabla_\mu \langle \overline{\phi}_l(\infty)\phi_k(0)\rangle
= {1\over (2\pi)^2}\int d^4x \langle {\cal
O}_\mu(x)\overline{\phi}_l(\infty) \phi_k(0)\rangle =0} since the
3-point function in the integrand vanishes\foot{This can be shown by following
similar logic as the one used to prove (4.3) below.}.

\newsec{Marginal deformations in ${\cal N}=2$ superconformal field theories}

We now focus on the main goal of this paper, the analysis of marginal
deformations of ${\cal N}=2$ superconformal field theories in four
dimensions. The important element is that in these theories the
marginal operators are descendants of chiral primaries \margoper.  As
a result, the variation of the chiral ring under marginal deformations
is tightly constrained.  In this section we present the main results
of our paper and leave some of the derivations for the next section
and the appendices.

\subsec{Holomorphy of the chiral ring and WDVV equations}
We start by analyzing the moduli dependence of 3-point functions of
chiral primaries.  We consider a deformation of the theory\foot{This
is only schematic, we do not assume that there is an ``action'' $S$ or
that the CFT has a Lagrangian description. By this notation we just
mean that we consider a deformation by a real linear combination of
the marginal operators ${\cal O}_m$ and $\overline{\cal O}_m$.}
$$
  S \rightarrow S + {\delta\lambda^m \over (2\pi)^2}\int d^4x \,{\cal O}_m(x)
+ {\delta \overline{\lambda}^m \over (2\pi)^2}\int d^4 x\, \overline{\cal O}_m (x)
$$
We denote by $\nabla_m,\overline{\nabla}_m$ the derivatives with
respect to holomorphic and antiholomorphic marginal operators
$$
(\nabla_m)_k^l = \delta_k^l\partial_m + A_{mk}^l
$$
where $A_{mk}^l$ is the connection.

First we show that the chiral ring coefficients are (covariantly)
holomorphic
\eqn\antiholz
{\overline{\nabla}_m C_{ij}^k = 0} which is a consequence\foot{From
the vanishing of the 4-point function follows that
$\overline{\nabla}_m C_{ij \overline{l}} = 0$. Then using that
$\overline{\nabla}_m g^{\overline{l}k}=0$ from \compat\ and $C_{ij}^k
= C_{ij\overline{l}}g^{\overline{l}k}$ we get \antiholz.}  of the
vanishing of the 4-point function
\eqn\vanthre{
  \langle \overline{\cal
  O}_m(x) \overline{\bphi}_l(\infty) \bphi_i(1) \bphi_j(0)\rangle =0}
  This a direct result of the superconformal Ward
  identity \basicwardb: we notice that the marginal operator is of the
  form $\overline{{\cal O}}_m
  = \overline{Q}^4\cdot \overline{\lphi}_m$.  Since all supercharges
  $\overline{Q}_{i\dot{a}}$ anticommute among themselves, we can pick
  one of them, call it ${\bf \overline{Q}}$, and pull it to the left
  of the other supercharges. So we write the marginal operator as
$$
\overline{\cal O}_m(x) =
\{{\bf \overline{Q}},\Lambda\}(x)
$$ 
for some operator $\Lambda$.  Now according to \basicwardb\ we move
${\bf \overline{Q}}$ to the other insertions.  We do not get any
contribution from $\phi_i(1)$ and $\phi_j(0)$ since they are chiral
primaries and are both annihilated by ${\bf \overline{Q}}$. We do not
get any contribution from $\overline{\phi}_l(\infty)$ because as we
explained there is no contribution from the point at infinity if the
operator is a superconformal primary.  So we find the desired
relation \vanthre. Notice that in this proof we have assumed that all
points are distinct $x\neq \{0,1,\infty\}$. According to the
discussion in section 3 we will never have to bring any two points on top
of each other and hence we do not have to worry about issues with contact terms.

Notice that this argument did not depend on the number of chiral
primaries inserted in the correlator, since we only needed to ``hide''
the antichiral operator $\overline{\phi}_l$ at infinity.  So it can be
used to show that more general ``extremal correlators'' are also
covariantly holomorphic
\eqn\extremalc{
\overline{\nabla}_m \langle \phi_1(x_1)...\phi_n(x_n) 
\overline{\phi}_{n+1}(z)\rangle =0}
where the $U(1)_R$ charges satisfy $R_{n+1} = \sum_{i=1}^n R_i$. If we
have two or more antichiral insertions then we cannot apply a similar
argument.

Now let us consider the holomorphic dependence of 3-point
functions of chiral primaries and show how to derive the four-dimensional analogue of the WDVV
equation \DijkgraafDJ,\DijkgraafQW,\WittenIG. In general we have
$ \langle {\cal
O}_m(x) \overline{\bphi}_k(\infty) \bphi_i(1) \bphi_j(0)\rangle \neq
0$ so the $C_{ij}^k$ can have non-trivial holomorphic dependence on
the coordinates $\lambda$ of ${\cal M}$
$$
\nabla_m C_{ij}^k \neq 0
$$
We can however prove the integrability condition
\eqn\integrability{
  \nabla_i C_{jk}^l = \nabla_j C_{ik}^l}
where we assumed that the indices $i,j$ correspond to chiral primaries of $\Delta=2$. 
This is a result of the following relation of 4-point functions
$$
 \langle {\cal O}_i(x) \overline{\bphi}_l(\infty) \bphi_j(1)
  \bphi_k(0)\rangle  = \langle {\cal O}_j(x) \overline{\bphi}_l(\infty) \bphi_i(1)
  \bphi_k(0)\rangle
$$
This can be derived in similar manner using the superconformal Ward
identities by moving the supercharges from ${\cal O}_i$ onto $\phi_j$,
which they transform into ${\cal O}_j$ (the contribution from $\phi_k$
can be killed according to the Ward identity in appendix A). Then by
acting with a conformal transformation we can switch the positions of
the operators $\phi_i$ and ${\cal O}_j$ leaving the other two
unchanged.

Next we consider the associativity of the chiral ring. We start with
the following 4-point function of chiral primaries
$$
G(x,y) = \langle\overline{\phi}_l(\infty) \phi_i(x) \phi_j(y) \phi_k(0) \rangle
$$
where we take $\phi_i,\phi_j$ to be Lorentz scalars. Since the OPE of
chiral primaries is nonsingular we have
$$
\lim_{x\rightarrow 0} \lim_{y\rightarrow 0} G(x,y) = C_{jk}^m C_{im\overline{l}}
$$
and
$$
\lim_{y\rightarrow 0} \lim_{x\rightarrow 0} G(x,y) = C_{ik}^m C_{jm\overline{l}}
$$
Associativity of the OPE (or crossing symmetry) implies that the two
limits should be the same. Multiplying with the inverse metric we get
the desired expression
\eqn\associativity{
C_{ik}^m C_{jm}^l = C_{jk}^m C_{im}^l}

\noindent Before we proceed, let us summarize the results we have found so far
$$
 \overline{\nabla}_j C_k
 =0,\qquad\quad\,\,\,\, \nabla_i \overline{C}_k = 0
$$ 
$$
\nabla_iC_j=\nabla_jC_i,\qquad
    \overline{\nabla}_i\overline{C}_j=\overline{\nabla}_j\overline{C}_i
$$
\eqn\btt{\quad\quad[C_i,C_j]=0,\qquad\quad [\overline{C}_i, \overline{C}_j]=0}
where the notation $C_i$ means that we think of it as a matrix
$(C_i)_k^l$.  These results are identical to those found in
two-dimensional superconformal field theories
%\foot{As in
%two-dimensional topological field theories, we could consider the
%topologically twisted ${\cal N}=2$ theory and define the ``topological
%metric'' by the 2-point function
%$\eta_{ij} \equiv \langle \phi_i \phi_j\rangle_{twisted}$ on ${\bf
%S}^4$ and the 3-point functions with all indices lowered
%$C_{ijk} \equiv
%\langle \phi_i\phi_j\phi_k\rangle_{twisted}$. Then the integrability
%equation \integrability\ implies that a scalar function ${\cal F}$
%exists (locally), such that $\eta_{ij} = \partial_i \partial_j {\cal
%F}(\lambda),\,C_{ijk} = \partial_i \partial_j \partial_k {\cal
%F}(\lambda)$ and the associativity condition \associativity\ can be
%written as a nonlinear differential equation
%$$
%{\partial^3 {\cal
% F} \over \partial \lambda_i \partial \lambda_k \partial \lambda_m}
%\left({\partial^2 {\cal F} \over
%\partial \lambda_m \partial \lambda_n}\right)^{-1}{\partial^3 {\cal F} \over \partial \lambda_n \partial \lambda_j
% \partial \lambda_l} = {\partial^3 {\cal
%F} \over \partial \lambda_j \partial \lambda_k \partial \lambda_m}\left({\partial^2
%{\cal F} \over
%\partial \lambda_m \partial \lambda_n}\right)^{-1}{\partial^3 {\cal F} \over \partial \lambda_n \partial \lambda_i
% \partial \lambda_l}
%$$
%This and other properties of the topologically twisted ${\cal N}=2$
%SCFTs in four dimensions will be discussed in \wipk.}.

The WDVV equations in ${\cal N}=2$ theories in four dimensions have
been discussed (in a different context)
in \BonelliQH,\MarshakovAE,\MarshakovNY,\MarshakovCR. These papers
focused on the prepotential of the low energy effective action on the
Coulomb branch of Seiberg-Witten theories, while we study the
dependence of correlators on the position on the moduli space
(i.e. parameter space) of the CFT.

\subsec{The $tt^*$ equations}

So far we have seen that the 3-point functions are (covariantly)
holomorphic functions of the coupling constants. Remarkably more
constraints can be found by computing the connection $A_{mk}^l$ on the
bundle of chiral primaries. From the point of view of conformal
perturbation theory this connection describes the operator mixing
under marginal deformations.  Its curvature is computed by \maincurv,
where the two marginal operators are now descendants of chiral
primaries.

As we show in the next section, in the case of ${\cal N}=2$
superconformal field theories, we can actually perform the double
integral of \maincurv\ and we find the following simple expressions
for the curvature\foot{In the first version of the paper the numerical coefficient
in front of the factor ${R\over c}$ in the last equation contained a mistake.}
$$
[\nabla_i,\nabla_j]_k^l=0
$$
$$
[\overline{\nabla}_i,\overline{\nabla}_j]_k^l=0$$
\eqn\maintt{  
        [\nabla_i,\overline{\nabla}_j]_k^l = -[C_i,\overline{C}_j]_k^l
+ g_{i\overline{j}}\delta_k^l\left(1+{ R\over4 c} \right)} where in
these equations the curvature operators are acting on the subspace of
chiral primaries of $U(1)_R$ charge $R$ which are labeled by the
indices $k,l$, while the indices $i,j$ denote the marginal operators
i.e. the corresponding chiral primaries of $\Delta=2$.  The constant
$c$ is the central charge of the conformal field theory, defined by
the 2-point function of the stress-energy tensor, $g_{i\overline{j}}$
is the 2-point function of chiral primaries of $\Delta=2$ and is
proportional to the metric on the moduli space $G_{i\overline{j}}=192
g_{i\overline{j}}$ (see
\normalr) and we use a condensed notation for 
$ [C_i,\overline{C}_j]_k^l =C_{ik}^r g_{r\overline{m}}
C^{* \overline{m}}_{\overline{j}\overline{p}}g^{\overline{p}l} -
g_{k\overline{r}}
C^{*\overline{r}}_{\overline{j}\overline{p}}g^{\overline{p}n}C^l_{in}
$.

These equations, whose proof is presented in section 5, are the main
results of our analysis. They were first discovered in the context of
two-dimensional ${\cal N}=(2,2)$ supersymmetric field theories and
called ``$tt^*$ equations'' \CecottiME. The first two equations show
that the bundles of chiral primaries ${\cal V}_p$ are holomorphic
vector bundles, while the last one allows us to compute the curvature
in terms of the chiral ring coefficients $C_{ij}^k$.

\subsec{Curvature of the supercurrents}

Before we continue we have to explain a
subtlety. The ${\cal N}=2$ superconformal algebra is invariant under a
$U(1)$ automorphism which rotates the left chiral supercharges as
$Q\rightarrow e^{i\theta}Q$ and the right chiral ones as
$\overline{Q}\rightarrow e^{-i\theta}
\overline{Q}$ leaving all bosonic generators unchanged\foot{Notice that
under this automorphism the superconformal partners rotate as $S\rightarrow e^{-i\theta} S\,,\,\,
\overline{S}\rightarrow e^{i \theta}\overline{S}$.}. 
This means that the phase of the supercharges is ambiguous and as we
will see there is nontrivial holonomy for it, under motion on the
moduli space. In other words the left chiral supercharges are sections
of a line bundle ${\cal L}$ over the moduli space and the right chiral
supercharges $\overline{Q}$ of its conjugate $\overline{\cal L}$.
               
The curvature of ${\cal L}$ is the same as that of the left chiral
supercurrents $G^i_{a\mu}$, since $Q_a^i = \int d^3 x G^i_{a0}$. So we
can use the general formula \maincurv\ for the curvature of the
supercurrents to compute the curvature of ${\cal L}$
$$
  F_{\mu\nu}^{\cal L} ={1\over (2\pi)^4} \int_{|x|\leq 1} d^4x \int_{|y|\leq 1}
d^4y \langle \overline{G}(\infty) {\cal O}_{[\mu}(x) {\cal
O}_{\nu]}(y) G(0)\rangle
$$
where for simplicity we did not write the spinor and flavor
indices. The curvature can be computed by similar methods as those
used in section 5 and after some work described in appendix D we find
the following\foot{In the first version of this paper the curvature of
the supercurrent was off by an overall numerical factor.}
\eqn\ccur{\eqalign{
  F_{ij}^{\cal L} &= 0\cr F_{\overline{i}\,\overline{j}}^{\cal L} &=
 0 \cr  F_{i\overline{j}}^{\cal L} &=- {1 \over 4 c}
 g_{i\overline{j}}}} So we learn that ${\cal L}$ is a holomorphic
 vector bundle whose curvature is proportional to the K\"ahler form of
 the moduli space. We remind that we are working in conventions where the 2-point function $g_{i\overline{j}}$ of chiral primaries of $\Delta=2$ is related to the 2-point function of marginal operators $G_{i\overline{j}}$ by $G_{i\overline{j}}=192\, g_{i\overline{j}}$.

\subsec{$tt^*$ equations as differential equations}

The set of equations \btt\ and \maintt\ can be made less abstract by
working in a ``holomorphic gauge'': from the second equation in
\maintt\ we see that it is possible to choose the basis of chiral 
primaries to depend on the coupling constants in such a way that
$A_{\overline{j}k}^l = 0$.  In that basis the covariant and ordinary
antiholomorphic derivatives coincide $\nabla_{\overline{j}} =
\partial_{\overline{j}}\,$ and equation \antiholz\ becomes
\eqn\holgaug{
  \overline{\partial}_{m} C_{ij}^k= 0} So in this basis the chiral
ring coefficients become ordinary holomorphic functions of the moduli
(and not just ``covariantly holomorphic''). If we know the global
structure of the moduli space ${\cal M}$ and the behavior of the
chiral ring coefficients in the weak-coupling limits it may be
possible to completely determine them for all values of the coupling.

To proceed we need to notice the following\foot{This subtlety
was not appreciated in the first version of the paper.}: if we want
the marginal operators to correspond to coordinate tangent vectors
(i.e. integrable and compatible with the complex structure of the
moduli space) the operators ${\cal O}_i$ must be holomorphic sections
of the tangent bundle. The marginal operators are related to the
chiral primaries of $R=4$ by
\eqn\margch{
\phi_i = Q^4\cdot {\cal O}_i
} As we mentioned above the supercharges are sections of a holomorphic
bundle ${\cal L}$. However in the standard conventions the
supercharges are not holomorphic sections, since we have taken their
2-point function to be constant over the moduli space (in order to
preserve the standard normalization $\{Q_a, \overline{Q}_{\dot{b}}\}=
2P_{a\dot{b}}$ of the algebra). Since the $Q$'s are not
holomorphic sections, we can see from \margch\ that in these standard
conventions it is not possible to simultaneously choose the $\phi_i$'s
and ${\cal O}_i$'s to be holomorphic sections. 
%% In order to overcome
%% this difficulty we have to relax the conventions \margch\ and consider
%% a more general normalization of the chiral primaries.
We can however relate the two different normalizations of the chiral primaries
as follows.

First let us introduce some notation. Let us call ${\cal K}$ the Kahler potential
for the metric on the moduli space i.e. $G_{i\overline{j}}=
\overline{\partial}_j \partial_i {\cal K}$. In our conventions the 2-point function of
chiral primaries of $\Delta=2$ is related to the metric on the moduli
space by $G_{i\overline{j}}=192 g_{i\overline{j}}$. So if we define $\widetilde{\cal K}=
{\cal K}/192$ we have $g_{i\overline{j}}=\overline{\partial}_j\partial_i \widetilde{\cal K}$. Let us call $\phi_i'$ the chiral primaries normalized so that they are holomorphic
sections and $\phi_i$ a different normalization related by
$$
\phi_i = e^{ { R\over 8 c} \widetilde{{\cal K}}} \phi_i'
$$
For the special case of chiral primaries of $R=4$ one can see that while
the $\phi_i'$'s are a holomorphic section,  the $\phi_i$'s are
those which give holomorphic sections after $Q^4$ acts on them,
which ensures that the marginal operators as defined by \margch\ are indeed
holomorphic sections. We have for the 2-point functions
\eqn\relmetr{
g_{i\overline{j}} = e^{ {R\over 4 c} \widetilde{{\cal K}}}  g_{i\overline{j}}'
}
In the holomorphic gauge, i.e. in the primed basis,     the
connection and curvature can be directly written in terms of the
metric 
\eqn\conhol{
  A_{ik}^l =  g^{\overline{m}l'}\,\partial_i g'_{k \overline{m}},\qquad A_{\overline{j}k}^l=0}
and hence 
$$[\nabla_i,\overline{\nabla}_j]_k^l = -\overline{\partial}_j A_{ik}^l =-
\overline{\partial}_j(g^{\overline{m}l'}\,\partial_i g'_{k \overline{m}}) 
$$
where we used the fact that the bundles are holomorphic and that the
metric and connection are hermitian and compatible\foot{From the
compatibility of the metric with the connection we have $\nabla_i
g_{k\overline{m}} =0$. Using that in the holomorphic gauge we have
$A_{\overline{j}k}^l = A_{j\overline{k}}^{\overline{l}}=0$ we can also
write the compatibility condition as $\partial_i g_{k\overline{m}} -
A_{ik}^l g_{l\overline{m}}=0$ and then \conhol\ follows.}. Using \relmetr\ this implies
$$
[\nabla_i,\overline{\nabla}_j]_k^l = -\overline{\partial}_j( g^{\overline{m}l}\,
\partial_i g_{k\overline{m}}
) + {R\over 4c } \overline{\partial}_j \partial_i \widetilde{{\cal K}}
\,\delta_k^l =-\overline{\partial}_j( g^{\overline{m}l}\,
\partial_i g_{k\overline{m}}
) +
{ R\over 4c} g_{i\overline{j}}\,\delta_k^l
$$
Putting everything together we find that the integrability
equations \btt\ become
\eqn\wdvvtwo{
  \partial_i C_{jk}^l - \partial_j C_{ik}^l = [ g^{-1}\partial_i g,
C_j]_k^l - [ g^{-1}\partial_j g, C_i]_k^l } and the last of the $tt^*$
equations \maintt\ can be written as
\eqn\ttdiff{\overline{\partial}_j( g^{\overline{m}l}\,
\partial_i g_{k\overline{m}}) = [C_i,\overline{C}_j]_k^l -
  g_{i\overline{j}}\delta_k^l} where it is important to emphasize that
  these equations hold in the choice of gauge (normalization)
  described above.

When written in the form \holgaug , \wdvvtwo\ and \ttdiff, the $tt^*$
equations can be considered as partial differential equations relating
the 2- and 3-point functions $g_{k\overline{l}}(\lambda)$ and
$C_{ij}^k(\lambda)$ of chiral primaries over the moduli space. One of
their applications is that they may allow us to determine the
Zamolodchikov metric $g_{k\overline{l}} (\lambda)$ (for the entire
tower of chiral primaries) for all values of the coupling.  The
important point is that while the 2-point functions
$g_{k\overline{l}}$ are not holomorphic, they can be computed as
solutions of these classical differential equations whose coefficients
are the holomorphic functions $C_{ij}^k(\lambda)$, which are in
general easier to determine.

\subsec{A flat connection}

Let us briefly comment on the  term
$g_{i\overline{j}} g_{k\overline{l}}\left(1 + { R \over 4c}\right)$
in the third equation of \maintt, which was absent in the original
derivation
\CecottiME\ of the 2-dimensional version of the $tt^*$ equations. There the connection for the Ramond
ground states of the topologically twisted theory was computed. In our
case we are computing the connection characterizing operator mixing in
a (four dimensional) conformal field theory. To check that in
our case we do need this extra term we notice that without it we would
have nontrivial holonomy (phase) for the identity operator in the CFT
(see also \deBoerSS), which seems unnatural.

In any case it is not difficult to relate the two results. Let us call
${\cal W}$ the holomorphic line bundle whose curvature is proportional
to the K\"ahler form on the moduli space
$$F_{ij}=F_{\overline{i}\overline{j}}=0\quad,\quad
F_{i\overline{j}} = g_{i\overline{j}}$$ And let us redefine each of
the bundles of chiral primaries ${\cal V}$ in the following way $V' =
{\cal W}^{-\left({R \over 4c}+1\right)}\otimes V$. Then the curvature
for the bundles $V'$ will be
$$
[\nabla'_i,\nabla'_j]_{k\overline{l}}=0
$$
$$
[\overline{\nabla}'_i,\overline{\nabla}'_j]_{k\overline{l}}=0
$$
\eqn\mainttb{  
[\nabla'_i,\overline{\nabla}'_j]_{k\overline{l}} =-
[C_i,\overline{C}_j]_{k\overline{l}} } which is the same as the one
discussed in \CecottiME.

Finally let us mention that the $tt^*$ equations, together with \btt,
are equivalent to the statement that one can define an improved
connection
$$
 D_i = \nabla'_i -C_{i}
$$
The improved connection is flat
$$
[D_i, D_j] = [\overline{D}_i,\overline{D}_j]= 
[{D}_i,\overline{D}_j] =0
$$
However this connection is not compatible with the Zamolodchikov
metric.

\newsec{Derivation of the four-dimensional $tt^*$ equations}

Now we show how the four-dimensional $tt^*$ equations can be derived
from the point of view of conformal perturbation theory. Let us start
with the first equation $[\nabla_i,\nabla_j]_k^l =0$. For this it is sufficient
to show that
\eqn\holhol{
\langle \overline{\bphi}_l(\infty) {\cal O}_{i}(x) {\cal O}_{j}(y) \bphi_k(0)\rangle
=0
}
We first act with a conformal transformation to interchange the points
at $\infty$ and $0$, so equivalently we want to prove the vanishing of
$\langle \bphi_k(\infty) {\cal O}_{i}(x) {\cal
O}_{j}(y) \overline{\bphi}_l(0)\rangle$. This is easy to prove as
before. We write ${\cal O}_i = \{{\bf Q,}\Lambda'\}$ and then
using \basicwardb\ we move ${\bf Q}$ to the other insertions. We find
that none of the terms contribute (where we also have to use that
$[{\bf Q},{\cal O}_j]=0$). So we find that the 4-point function
vanishes. Similarly we show that
$[\overline{\nabla}_i,\overline{\nabla}_j]_k^l=0$.

We will now show how to derive the third of the $tt^*$ equations \maintt. For
that we need to compute the following integrated 4-point function
$$
(F_{i\overline{j}})_{k\overline{l}} \equiv
[\nabla_i,\overline{\nabla}_j]_{k\overline{l}}={1\over (2\pi)^4}
\int_{|x|\leq 1} d^4x \int_{|y|\leq 1} d^4y \langle \overline{\bphi_l}(\infty)
 {\cal O}_{[i}(x) \overline{\cal O}_{j]}(y)\bphi_k(0)\rangle
$$
The superconformal Ward identities of the ${\cal N}=2$ CFT allow us to
write
$$
\langle\overline{\bphi}_l(\infty)  {\cal O}_{i}(x) \overline{\cal O}_{j}(y)
 \bphi_k(0)\rangle
  =\Boxx \Boxx \langle\overline{\bphi}_l(\infty) \lphi_{i}(x) \overline{\lphi}_{j}(y) \bphi_k(0)\rangle
$$
where $\Boxx$ is the four-dimensional Laplacian with respect to $x$. A
quick explanation of this identity is as follows: we have
$\overline{{\cal O}}_j = \overline{Q}^4 \cdot \overline{\lphi}_j$. We
want to move the $\overline{Q}$'s away from the operator
$\overline{\lphi}_j(y)$ using \basicwardb. The operator $\bphi_k$ is
annihilated by the $\overline{Q}$'s being an antichiral primary and $\overline{\phi}_l$
does not contribute according to \basicwardb, so
the only contribution will be when the $\overline{Q}$'s hit the ${\cal
O}_i$. Then we end up with something of the form
$ \overline{Q}^4 \cdot Q^4\cdot\lphi_i(x)$. From the supersymmetry
algebra $\{Q_a^i,\overline{Q}_{j,\dot{b}}\} = 2 \delta^i_j
P_{a\dot{b}}$ we see that we get derivatives $\partial_x$ when we
anticommute the supercharges. The only Lorentz invariant combination
we can construct out of four $\partial_x$'s is $\nabla^2_x\nabla^2_x$.

So we have
\eqn\aaa{\eqalign{
(F_{i\overline{j}})_{k\overline{l}} =&{1\over (2\pi)^4}
\int_{|x|\leq 1} d^4x \int_{|y|\leq 1}
  d^4y\cr &\left(
\Boxx \Boxx  \langle\overline{\bphi}_l(\infty)
  \lphi_{i}(x) \overline{\lphi}_{j}(y) \bphi_k(0)\rangle
-\Boxy \Boxy \langle\overline{\bphi}_l(\infty) \lphi_{i}(y) \overline{\lphi}_{j}(x) \bphi_k(0)\rangle\right)}}
Now we use the following conformal Ward identity\foot{This identity is
true for scalar conformal primaries $\lphi_i,\overline{\lphi}_j$ of
dimension $\Delta=2$ and for scalar conformal primaries
$\bphi_k,\overline{\bphi}_l$ of any (but same) conformal dimension. If
$\bphi_k,\overline{\bphi}_l$ are primaries with nonzero spin, then the
identity is still true after integrating over the angular directions
of the variable $x$. It does not depend on supersymmetry. The easiest way to check it is
to consider the double OPE of both sides taking say $x$ close to $\infty$ and $y$ close to $0$.}
$$
  \Boxx \Boxx \langle\overline{\bphi}_l(\infty) \lphi_{i}(x) \overline{\lphi}_{j}(y) \bphi_k(0)\rangle=
\Boxy \Boxy\left({ |y|^4 \over |x|^4} 
 \langle\overline{\bphi}_l(\infty)
  \lphi_{i}(x) \overline{\lphi}_{j}(y) \bphi_k(0)\rangle\right)
$$
Using this result and the OPEs between chiral and antichiral primaries
(appendix E) we can see that the integrand in \aaa\ is smooth as
$y\rightarrow 0$, so doing the $y$-integration
\eqn\aab{\eqalign{
(F_{i\overline{j}})_{k\overline{l}} =&{1\over (2\pi)^4}\int_{|x|\leq 1}
d^4x \int_{|y|=1} d\Omega_3^y |y|^2 (y\cdot \partial_y) \cr &
\left(\Boxy{ |y|^4 \over |x|^4} 
 \langle\overline{\bphi}_l(\infty)
  \lphi_{i}(x) \overline{\lphi}_{j}(y) \bphi_k(0)\rangle-
\Boxy \langle\overline{\bphi}_l(\infty)
  \lphi_{i}(y) \overline{\lphi}_{j}(x) \bphi_k(0)\rangle\right)}} Now
we use two more conformal Ward identities\foot{Same is true as in the
previous footnote.}
\eqn\aag{
  \Boxy \langle\overline{\bphi}_l(\infty) \lphi_{i}(y) \overline{\lphi}_{j}(x) \bphi_k(0)\rangle=
\Boxx \left( {|x|^2 \over |y|^2} \langle\overline{\bphi}_l(\infty)
  \lphi_{i}(y) \overline{\lphi}_{j}(x) \bphi_k(0)\rangle\right)}
and 
\eqn\aac{\eqalign{
&\int_{|x|=const} d\Omega_3^x \, \Boxy (
  |y|^4\langle\overline{\bphi}_l(\infty) \lphi_{i}(x) \overline{\lphi}_{j}(y) \bphi_k(0)\rangle)=
\cr & =\int_{|x|=const} d\Omega_3^x  \, |y|^2(x \cdot \partial_x) \left(|x|^2 (x \cdot \partial_x) \left({1\over |x|^2} \langle\overline{\bphi}_l(\infty)
  \lphi_{i}(x) \overline{\lphi}_{j}(y) \bphi_k(0)\rangle\right)\right)}}
\fig{The colored dotted lines denote surface integrals of chiral primary operators and their conjugates.}
{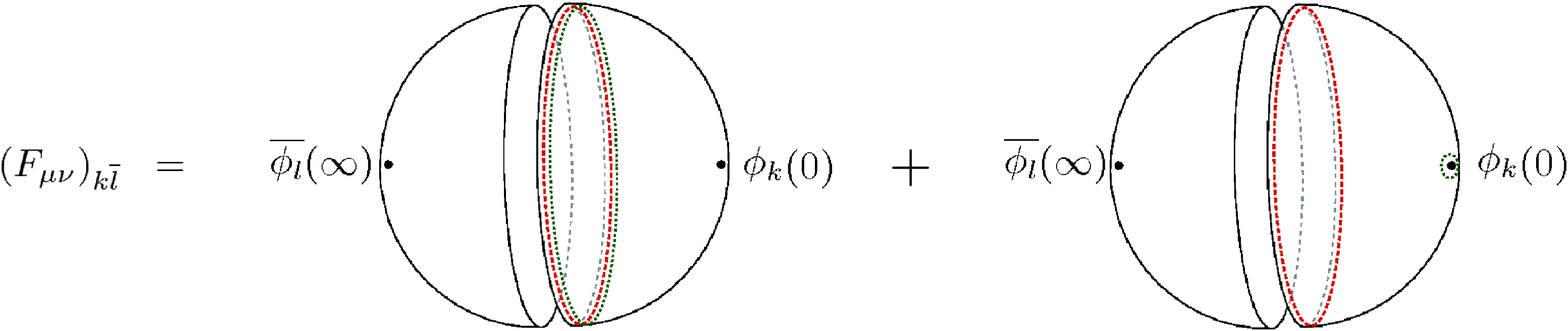}{6.truein}
\figlabel{\curvatureb}
Substituting \aag\ and \aac\ into \aab\ and doing the $x$ integral we
find
\eqn\aae{\eqalign{
  (F_{i\overline{j}})_{k\overline{l}} =&\quad {1\over (2\pi)^4}\lim_{r\rightarrow 1^-}
\int_{|x|=r}d\Omega_3^x  \int_{|y|=1} d\Omega_3^y |x|^2|y|^2 (y\cdot \partial_y)(x\cdot \partial_x) \cr
& \qquad\left( { |y|^2\over
  |x|^2} \langle\overline{\bphi}_l(\infty) \lphi_{i}(x) \overline{\lphi}_{j}(y) \bphi_k(0)\rangle
  - {|x|^2 \over
  |y|^2}\langle\overline{\bphi}_l(\infty) \lphi_{i}(y) \overline{\lphi}_{j}(x) \bphi_k(0)\rangle\right)\cr
  & -{1\over (2\pi)^4} \lim_{r\rightarrow 0} \int_{|x|=r}d\Omega_3^x \int_{|y|=1}
  d\Omega_3^y |x|^2|y|^2 (y\cdot \partial_y)(x\cdot \partial_x) \cr
  & \qquad
\left( { |y|^2\over |x|^2} \langle\overline{\bphi}_l(\infty)
  \lphi_{i}(x) \overline{\lphi}_{j}(y) \bphi_k(0)\rangle -
 {|x|^2 \over
 |y|^2}\langle\overline{\bphi}_l(\infty) \lphi_{i}(y) \overline{\lphi}_{j}(x) \bphi_k(0)\rangle\right)}}
 which is schematically depicted in figure \curvatureb. The
 contribution from the limit $r\rightarrow 0$ can be easily evaluated
 by considering the OPE between the operator at $x$ and the one at
 $0$. We have
\eqn\aaf{\eqalign{
& \lphi_i(x) \bphi_k(0) = C_{ik}^m
\, \bphi_m(0)+...\cr
& \overline{\lphi}_j(x) \bphi_k(0) =
g_{k\overline{r}}C^{*\overline{r}}_{
\overline{j}\overline{p}}
    g^{\overline{p}n}{\bphi_n(0) \over |x|^4}+...  }} where we did not
include any higher order terms since they do not contribute to the
integral in the limit $r\rightarrow 0$. Notice that the first
expression is the chiral ring OPE while the second is explained in
appendix E. So the contribution from this limit is
$$
   g_{k\overline{r}}
C^{*\overline{r}}_{\overline{j}\overline{p}}g^{\overline{p}n}C^m_{in}
g_{m\overline{l}} - C_{ik}^r g_{r\overline{m}}
C^{* \overline{m}}_{\overline{j}\overline{l}} = -
[C_i,\overline{C}_j]_{k\overline{l}}
$$
Then for the contribution to \aae\ from the $r\rightarrow 1$ limit, we
notice that there are many cancellations between the integrands at
points related by $x\leftrightarrow y$ as $r\rightarrow 1$, and the
only possible nonzero contribution is from the vicinity of the region
$x=y$. There we are allowed to use the OPE between $\lphi_i$ and
$\overline{\lphi}_j$. As we explain in appendix C, the only terms
which do contribute are the identity operator and operators in the
supermultiplet containing the stress-energy tensor. The contribution
to the OPE of these operators is fixed by Ward identities so finally
we find the following result
$$
    g_{i\overline{j}}g_{k\overline{l}}\left(1+{ R\over 4 c} \right)
$$
Putting everything together we have
$$
(F_{i \overline{j}})_{k\overline{l}}
=-[C_i, \overline{C}_j]_{k\overline{l}} + g_{i\overline{j}}g_{k\overline{l}}
\left(1+{ R\over 4 c} \right)
$$

\newsec{The Zamolodchikov metric on the moduli space}

The previous results allow us to find constraints for the geometry of
the moduli space of general ${\cal N}=2$ superconformal field
theories. In this section we show that the Zamolodchikov metric
satisfies constraints similar to those of ``special geometry''. The
marginal operators are of the form
\eqn\margopp{
{\cal O}_i = Q^4\cdot \lphi_i \qquad ,\qquad \overline{\cal O}_j
= \overline{Q}^4\cdot \overline{\lphi}_j} From this we see that the
holonomy of the marginal operators is related to that of the chiral
primaries of $\Delta=2$. More specifically the marginal operators ${\cal O}_i$ are
sections of the holomorphic tangent bundle ${\cal TM}$.  From
expression \margopp\ we see that we have
$$
{\cal TM} = {\cal L}^4 \otimes {\cal V}
$$
where ${\cal L}$ is the $U(1)$ bundle of the supercharges and ${\cal
V}$ is the vector bundle of the chiral primaries $\lphi_i$ of
$\Delta=2, R=4$. The Riemann tensor on the moduli space is determined
by the curvature of ${\cal TM}$, so it will be the sum of the
curvatures of the bundles ${\cal V}$ and that of ${\cal L}^4$. The
curvature of ${\cal V}$ is given  by the $tt^*$ equations \maintt\
for the chiral primaries with $R=4$ and that of ${\cal L}^4$ is four
times that of ${\cal L}$ given in
\ccur. Putting everything together we find the following equation for
the Riemann tensor on the moduli space
\foot{In the first version of this paper this formula contained an incorrect factor which has been fixed now.}
\eqn\zriem{
  R_{i\overline{j}k}^l =- C_{ik}^M
 g_{M\overline{N}}C^{*,\overline{N}}_{\overline{j}\overline{q}}g^{\overline{q}l}
 +g_{k\overline{j}}\delta_i^l +g_{i\overline{j}}\delta_k^l } where $C_{ij}^M$ are the chiral ring
 coefficients between chiral primaries $\lphi_i,\lphi_j$ of $\Delta
 =2$ and $\bphi_M$ of $\Delta =4$ and $g_{i\overline{j}}$ is
 proportional to the metric on the moduli space $G_{i\overline{j}}$ as $G_{i\overline{j}}=192 g_{i\overline{j}}$
 (see
\normalr).

These are partial differential equations for the Zamolodchikov metric
$G_{i\overline{j}}$ on ${\cal M}$, with coefficients given by the
holomorphic correlators $C_{ij}^M$ and their complex conjugates. These
equations are generalizations of special geometry and similar to those
describing the Weil-Petersson metric on the moduli space of Calabi-Yau
n-folds or those of ${\cal N}=(2,2)$ SCFTs in two dimensions with
arbitrary central
charge \deWitPK,\StromingerPD,\PeriwalMX,\BershadskyCX,\GreeneVM.

\newsec{${\cal N}=1$ theories}

Let us now explain which of our previous results are true for theories
with ${\cal N}=1$ superconformal invariance in four dimensions.  The
${\cal N}=1$ SCFT has a $U(1)_R$ R-symmetry. The standard
normalization of the $U(1)_R$ in ${\cal N}=1$ theories is such that
the unitarity bound is $\Delta \geq {3\over 2} R$. Superconformal
primary operators saturating this bound are chiral primaries and are
annihilated by the supercharges of right chirality
$\overline{Q}_{\dot{a}}$. We have the structure of a chiral ring as in
${\cal N}=2$ theories and we use the same notation for the 2- and
3-point functions $g_{i\overline{j}}$ and $C_{ij\overline{k}}$ and the
chiral ring coefficients $C_{ij}^k$.

In ${\cal N}=1$ SCFTs marginal deformations preserving supersymmetry
are descendants of chiral primaries $\Phi_i$ of $\Delta=3$ and again
they split into holomorphic and antiholomorphic ones
\eqn\margoper{
{\cal O}_i = Q^2\cdot \lphi_i \qquad ,\qquad \overline{\cal O}_j
= \overline{Q}^2\cdot \overline{\lphi}_j}

All of the results of section 4.1 are also true for ${\cal N}=1$ theories. In
particular the 3-point functions of chiral primaries satisfy the following set of 
equations
$$
 \overline{\nabla}_j C_k
 =0,\qquad\quad\,\,\,\, \nabla_i \overline{C}_k = 0
$$ 
$$
\nabla_iC_j=\nabla_jC_i,\qquad
    \overline{\nabla}_i\overline{C}_j=\overline{\nabla}_j\overline{C}_i
$$
\eqn\btt{\quad\quad[C_i,C_j]=0,\qquad\quad [\overline{C}_i, \overline{C}_j]=0}
where the notation $C_i$ means that we think of it as a matrix
$(C_i)_k^l$. We can also show that the following relations are true
$$
[\nabla_i,\nabla_j]_k^l=0
$$
$$
[\overline{\nabla}_i,\overline{\nabla}_j]_k^l=0
$$ 
The last equation implies that we can pick a holomorphic gauge as
before and then we have the following equations
\eqn\holgaug{
  \partial_{\overline{m}} C_{ij}^k= 0}
\eqn\wdvvtwo{
  \partial_i C_{jk}^l - \partial_j C_{ik}^l = [ g^{-1}\partial_i g,
C_j]_k^l - [ g^{-1}\partial_j g, C_i]_k^l }
We can also show that ``extremal'' correlators have to be holomorphic
\eqn\extremalc{
\overline{\nabla}_m \langle \phi_1(x_1)...\phi_n(x_n) 
\overline{\phi}_{n+1}(z)\rangle =0}
It would be very interesting to see whether the third of the $tt^*$
equations \maintt\ can also be derived for ${\cal N}=1$ SCFTs, possibly
with some modifications. We hope
to report on this in the future \wipk.

\newsec{${\cal N}=4$ theories}

Now we discuss how the previous analysis applies to theories with
${\cal N}=4$ superconformal invariance. Certain aspects of the
connection on the space of operators for ${\cal N}=4$ theories have
been discussed in \PetkouFV, \BasuNT.  An ${\cal N}=4$ theory can also
be viewed as an ${\cal N}=2$ theory and has several kinds of short
supermultiplets \MinwallaKA,\DolanZH. Superconformal multiplets of the
${\cal N}=4$ algebra are labeled by the conformal dimension $\Delta$,
the spin $(j,\overline{j})$ and by their $SO(6)_R$ R-symmetry
representation.

\subsec{The moduli space for ${\cal N}=4$ gauge theories}
All examples of ${\cal N}=4$ superconformal field theories that we
know are realized as Lagrangian gauge theories with a product gauge
group $\prod_{i=1}^n G_i$, each with a coupling constant
$\tau_i$. Then the moduli space is the direct product of the
individual moduli spaces, at least locally. So for simplicity we can
focus on the case where we only have a single gauge group.

The marginal operator in that case is a descendant of fields in a
short multiplet whose superconformal primaries are scalar fields
$\lphi$ with conformal dimension $\Delta=2$ and transforming into the
rank two symmetric traceless representation of $SO(6)$\foot{The gauge
invariant operators $\lphi$ should not be confused with the
adjoint-valued elementary scalar fields $X_i$ $i=1,...6$ of the ${\cal
N}=4$.  Schematically we have $\lphi \sim {\rm Tr} (X_{(i} X_{j)})$,
where we did not write down the $SO(6)_R$ indices of $\lphi$ and
parentheses around the indices denote symmetrization and subtraction of the trace.}. It is
not hard to show that the $SO(6)_R$ quantum numbers of fields do not
change under motion on the moduli space. So the holonomy on the
multiplet $\lphi$ must commute with $SO(6)$ which implies that the
holonomy is actually trivial.  So the multiplet $\lphi$ does not
receive any holonomy under motion on the moduli space.

This means that the curvature of the tangent bundle is simply given by
the curvature of the bundle ${\cal L}$ of the
supercharges\foot{Similar to the ${\cal N}=2$ superconformal algebra
the ${\cal N}=4$ algebra has an outer automorphism under which
$Q_a^i \rightarrow e^{i\theta}
Q_a^i,\, \overline{Q}_{i,\dot{a}}\rightarrow e^{-i\theta }
\overline{Q}_{i,\dot{a}}$.} which being proportional to the
K\"ahler form, is covariantly constant. Hence the moduli space is
locally the homogeneous hyperbolic\foot{From \ccur\ we can see that
the scalar curvature is negative.} space 
\eqn\lcoset{
  {SO(1,2) \over SO(2)}} It is standard to parametrize it in terms of
the complexified gauge coupling
$$
  \tau = {\theta \over 2\pi} + i {4\pi \over g_{YM}^2}
$$
with the metric
$$
  ds^2 \sim c \, {d\tau d\overline{\tau} \over {\rm Im}\tau^2}
$$
The global structure depends on the action of the S-duality group and
cannot be determined by our local analysis. We see from \ccur\ that
the overall scale of the metric is proportional to the central charge
$c$. This can be independently checked by a weak coupling computation
of the Zamolodchikov metric in the ${\cal N}=4$ SYM. In the case of
the ${\cal N}=4$ SYM with gauge group $SU(N)$ this metric agrees in
the large $N$ limit with the metric on the moduli space of the
axion-dilaton of IIB supergravity in $AdS_5\times S^5$.

If we have $n$ gauge group factors we will have $n$ corresponding
short multiplets $\lphi_i$ which give marginal operators and then the
moduli space is (locally) the product of spaces of the form
\lcoset. The statement that the gauge group is a direct product is related to the
fact that the $\lphi_i$'s do not mix among themselves under motion on 
the moduli space. 

\subsec{Exotic ${\cal N}=4$ theories?}

In this speculative paragraph let us mention that it is a logical
possibility that exotic ${\cal N}=4$ superconformal field theories may
exist, not realizable as weakly coupled Lagrangian theories anywhere
on their moduli space, where the space of coupling constants is not
just a direct product of ``gauge couplings''.  That would correspond
to an ${\cal N}=4$ theory with $n$ short supermultiplets of
$\Delta=2$, which contain marginal operators in their descendants, and
which do mix nontrivially under motion on the moduli space. How
complicated can this mixing be?  Since the mixing has to commute with
the $SO(6)$ R-symmetry we conclude that it can be at most an $SO(n)$
transformation.

The moduli space of such a theory would have real dimension $2n$. The
most general holonomy of a 2n-dimensional manifold is $SO(2n)$. On the
other hand we know that the holonomy of the tangent bundle is the
product of the $SO(2)$ holonomy of the supercharges with the $SO(n)$
of the $n$ short multiplets, so $SO(2)\otimes SO(n)$ in total. Then
the moduli space is a manifold of reduced holonomy.  Berger's
classification of manifolds with reduced holonomy then fixes the
metric on the moduli space and we learn that the moduli space is
locally of the form
$$
{SO(n,2) \over SO(n) \otimes SO(2)}
$$
This is the most general moduli space consistent with ${\cal N}=4$
superconformal invariance in four dimensions\foot{The argument we
followed is analogous to the one in \CecottiKZ\ generalizing the
results of \SeibergPF\ for the moduli space of two-dimensional ${\cal
N}=(4,4)$ theories.}.  As far as we know, there are no indications for
the existence of any exotic ${\cal N}=4$ theory of this type.

\subsec{Chiral primaries for ${\cal N}=4$ theories}

For chiral primaries of the ${\cal N}=4$ one can show that the chiral
ring coefficients are covariantly constant along {\it all} directions
of the moduli space. Then taking the covariant derivatives of both
sides of the $tt^*$ equations we get that the curvature of the bundles
of chiral primaries are covariantly constant
$$
\nabla F =0
$$
In the previous section we found that the moduli space of ${\cal N}=4$
theories is a homogeneous space. Bundles of covariantly constant
curvature over homogeneous spaces are called homogeneous bundles and
their structure is completely determined by group theory \Helgason, in
analogy with the two-dimensional story in \deBoerSS. For Lagrangian
${\cal N}=4$ theories this implies that the bundles of chiral
primaries are (locally) flat. For ``exotic'' ${\cal N}=4$
theories, if they exist, we would have nontrivial homogeneous bundles
labeled by representations of $SO(n)$. These issues will be discussed
in more detail in \wipk.

\newsec{A comment on the moduli space of vacua}

In this paper the term ``moduli space'' has been used to refer to the
parameter space of the conformal field theory. This has to be
distinguished from the moduli space of vacua. The first parametrizes a
family of conformal field theories which are continuously connected,
while the second is a family of vacua of a given and fixed conformal
field theory. The first is a characteristic of the quantum field
theory in the UV and refers to a tuning of the parameters in its
Lagrangian, while the second is an IR concept, which is characterized
by vacuum expectation values of certain operators.

Notice that in two-dimensional conformal field theory this distinction
is not usually made because in two dimensions there cannot be moduli
spaces of vacua due to infrared divergences. The situation is
different in higher dimensional conformal field theories where it is
possible for operators to get nonzero vevs.

It is also natural to consider the total space, the space of all
possible vacua of all continuously connected conformal field
theories. This has the structure of a bundle where the base is the
moduli space of conformal field theories and the fiber is the space of
vacua for the given theory \ArgyresBN. Perhaps it would be interesting
to explore the topology and geometry of this total space in more
detail.

In many Lagrangian ${\cal N}=2$ gauge theories this total space is
nicely described by the Seiberg-Witten curves \SeibergRS,\SeibergAJ,
which in the case of finite theories depend on the marginal coupling
$\tau$ and the vevs of the gauge invariant operators which parametrize
the position on the Coulomb branch. In Lagrangian theories we look for
vacua by minimizing a potential and computing the vevs of gauge
invariant operators on the space of solutions.  To understand the
similar structure for more general superconformal field theories,
without Lagrangian description, it would be necessary to work with the
concept of a ``vacuum'' in more abstract conformal field theory
language. In this framework we imagine that we are only given the list
of the primary (gauge invariant) operators ${\cal O}_i$ and their
conformal dimensions $\Delta_i$, as well as the 3-point functions
$C_{ij}^k$. In a nontrivial (i.e. non-conformaly invariant) vacuum the
primaries can get nonzero vevs $\langle{\cal O}_i\rangle = A_i
M^{\Delta_i}$, where $A_i$ are dimensionless numbers and $M$ is a mass
scale which characterizes the breaking of the scale invariance in the
vacuum. Only some combinations of vevs are allowed by the dynamics of
the theory and the set of such allowed combinations $\{A_i\}$ is the
moduli space of vacua of the CFT. It would be interesting to formulate
these conditions in terms of the CFT data $\Delta_i$ and $C_{ij}^k$.

\newsec{Discussions}

We derived some new exact results for the chiral ring of general
${\cal N}=2$ superconformal field theories and found structures
similar to those encountered in two-dimensional ${\cal N}=(2,2)$
theories.  Our analysis was based on conformal perturbation theory. It
would be interesting to study whether our equations can be derived in
an alternative way by considering the topologically twisted ${\cal
N}=2$ theory \WittenZE\ and following similar arguments as
in \CecottiME. Such an approach might also give results about massive
deformations of ${\cal N}=2$ superconformal field theories, which
could be useful for applications to theories with less
supersymmetry. It might also be interesting to explore the
topologically twisted ${\cal N}=2$ superconformal field theories on
more general four-manifolds. For such an approach the analysis
of \PestunRZ\ might be useful and also
\ShapereZF\ for recent discussions.

It would be instructive to apply our somewhat abstract analysis to
specific examples, such as the finite ${\cal N}=2\,SU(N)$ gauge
theories with ${N_f =2N}$ flavors in the fundamental. In these
theories the chiral ring is generated by chiral primaries of the form
${\rm Tr}(\Phi^k)$ where $\Phi$ is the adjoint scalar field and the
moduli space is parametrized by the gauge coupling constant
$\tau={\theta \over 2\pi} +i{4\pi \over g_{YM}^2}$. It is possible to
perform perturbative computations in the $\tau \rightarrow i \infty$
limit and then we can use the holomorphy of the chiral ring
coefficients and the $tt^*$ equations to extend the computation in the
interior of the moduli space.  In particular it would be interesting
to try to connect the $\tau\rightarrow i\infty$ point with the S-dual
``infinitely-strongly coupled'' limit $\tau\rightarrow 1$ discussed
in \ArgyresCN.

More generally it would be interesting to analyze the rich class of
${\cal N}=2$ superconformal field theories studied by Gaiotto
in \GaiottoWE. These theories have more complicated moduli spaces
related to those of punctured Riemann surfaces. Based on this, it was
pointed out in \GaiottoGZ\ that the Zamolodchikov metric on the moduli
spaces of such conformal field theories is the Weil-Petersson metric
on the Teichmuller space, which is consistent with the equations that
we found in section 6. The chiral primary operators of the
four-dimensional theory are related to holomorphic differentials on
the Riemann surface on which the M5 branes are wrapped. Thus the
connection for the chiral primaries that we computed is related to the
Gauss-Manin connection for the cohomology elements of the Riemann
surface under variation of its complex structure moduli. It would be
nice to make these statements more precise.

It might also be useful to explore our results from the six-dimensional
point of view. A six-dimensional superconformal field theory can be
either wrapped on a Riemann surface giving an ${\cal N}=2$ theory in
four dimensions, or on a 4-manifold resulting into a SCFT in 2 dimensions.
Both theories obey the $tt^*$ equations and it would be nice to understand
relations between the two and possible connections with \AldayAQ,\AldayFS.

Another interesting direction is clarifying the bulk interpretation of
our results in cases of ${\cal N}=2$ theories with AdS duals. The
moduli space of the conformal field theory is mapped to the moduli
space of the AdS compactification of supergravity/string theory. The
chiral primaries are related to certain Kaluza-Klein modes in the
internal manifold, or possibly other supersymmetric probes such as
giant gravitons.

It is worth trying to extend our results in ${\cal N}=1$
superconformal field theories in four dimensions. Upon dimensional
reduction on a two-torus they lead to two-dimensional ${\cal N}=(2,2)$
supersymmetric theories (not necessarily conformal). Hence one might
hope that some degree of control exists even in the parent
four-dimensional theory, see also \IntriligatorJR.  From the AdS/CFT
point of view the structure of the moduli space of ${\cal N}=1$
superconformal field theories was briefly discussed in \TachikawaTQ\
in relation to a-maximization \IntriligatorJJ. It might be possible to
explore these questions using methods similar to the ones developed in
this paper.

Our analysis has been local on the moduli space. It would be
interesting to analyze certain global issues such as the action of the
S-duality group on correlators and possible monodromies for the chiral
primaries around non-contractible cycles. Of interest would also be to
analyze what happens at points where chiral primaries combine into
long multiplets and lift \KinneyEJ.  In a sense these seem like the
SCFT analogue\foot{Here we are talking about joining and splitting of
BPS multiplets of the superconformal group, i.e. of local operators,
as a function of the position on the moduli space of the CFT and not
about the joining and splitting of BPS dyons as a function of the
position on the moduli space of vacua. Unlike what
happens for BPS dyons, the joining and splitting of BPS local
operators happens along surfaces of codimension larger than one, so
the term ``wall'' in this case is not so accurate. It would be
useful to understand the rules of joining and splitting of such BPS states
and possible relations with \KS,\GaiottoCD,\CecottiUF. } of ``walls of
marginal stability''.

It would be nice to check that the renormalization method that we
adopted to implement conformal perturbation theory is actually consistent 
with all axioms of conformal field theory and with supersymmetry, especially
at higher orders in the perturbation.

Finally let us mention that the connection of operators that we have
computed can be interpreted as Berry's phase \BerryJV\ (and its non-abelian
generalization \WilczekDH) for a class of states as we now explain. If we define
the conformal field theory on ${\bf S^3}\times{\rm time}$ then it can
be thought of as a quantum system whose Hilbert space is isomorphic to
the set of local operators on the plane, via the state-operator map.
These states fall into unitary representations of the superconformal
algebra.  A special class of these states, the ones that are dual to
the chiral primary operators, belong to shorter representations. If we
adiabatically vary the coupling constants of the theory, then states
are subject to the phenomenon known as geometric or Berry's phase. The
$tt^*$ equations compute the Berry phase for the chiral states. In the
AdS bulk this would be the Berry phase of certain quantum states under
adiabatic variation of the supergravity moduli.  See also \deBoerQE\
and \PedderWP,\PedderFF,\PedderJE,\SonnerFI.
\bigskip
\bigskip
\bigskip
 \centerline{\bf Acknowledgments}

\noindent We would like to thank L. Hollands,  E. Kiritsis, 
M. Shigemori, M. Taylor, D. Tong, C. Vafa, E. Witten and especially
J. de Boer, K. Skenderis and E. Verlinde for useful discussions and
comments on the manuscript. We thank the University of Crete for
hospitality during the completion of this work.

\appendix{A}{Superconformal Ward identities}

We will show how to derive the Ward identities \basicward\
and \basicwardb\ in more detail.  As a warm-up we review the
derivation of the Ward identity for a conserved bosonic current. Of
course this is a standard result, but we want to derive it from a more
abstract conformal field theory point of view, without any reference
to an action or the Noether procedure. Our starting point is that the
CFT has a spin $(1/2,1/2)$ operator $J^\mu$ of $\Delta=3$. Using the
conformal algebra we can compute the norm of the first descendant
$\partial_\mu J^\mu$ and we find that it vanishes, which implies the
operator equation $\partial_\mu J^\mu = 0$, so $J^\mu$ is a conserved
current. We define the corresponding conserved charge by $R = \int
d^3x\, J^0$.  For any local operator $\varphi$ we have the following
OPE
$$
  J^\mu(x) \varphi(0) = ...+ {1\over 2\pi^2} {x^\mu \over |x|^4} [R,\varphi](0)+...
$$
with no other terms of the form ${x^\mu \over |x|^4}$. This can be
written as
\eqn\littlec{
[R,\varphi](x) = \oint_x dS_\mu \cdot J^\mu(z) \varphi(x) } where we
compute the integral over a small sphere\foot{We compute the integral
on a sphere of radius $\epsilon$ and then take the limit
$\epsilon\rightarrow 0$.} surrounding the point $x$. Now consider the
general correlator
$$
 V^\mu(z) = \langle J^\mu (z) \varphi_1(x_1)...\varphi_n(x_n)\rangle
$$
where we assume that all points are distinct. If we keep $x_k$ fixed
and think of $V$ as a function of $z$ then it is vector field defined
on ${ \cal D} = {\bf R}^4-\{x_1,...,x_n\}$, which is divergenceless $
\partial_\mu V^\mu =0$, since $\partial_\mu J^\mu=0$. Then
\eqn\gaussin{
0 = \int_{\cal D}\partial_\mu V^\mu =
\int_{\partial{\cal D}} dS_\mu\cdot V^\mu =  \oint_\infty dS_\mu \cdot V^\mu(z) - \sum_i \oint_{x_i} dS_\mu \cdot V^\mu(z)
} Since $J^\mu$ has conformal dimension $\Delta=3$ we know that
$V^\mu(z)$ falls off at least as ${1\over |z|^6}$ at large $z$, so
there is no contribution from the sphere at infinity in \gaussin\ and
we only get contributions from the punctures $\{x_1,...,x_n\}$. Using
\littlec\ we find
\eqn\wardcur{\sum_k \langle \varphi_1(x_1)...[R,\varphi_k](x_k)...\varphi_n(x_n)\rangle =0}
which is the desired Ward identity expressing charge conservation.

Now we will do the same thing for the supercurrent. The left chiral
supercurrent $G_a^{i\mu}$ is a conformal primary operator of dimension
$\Delta = 7/2$ transforming in the $(1,1/2)$ representation of the
Lorentz group (indices $a,\mu$) and in the $I=1/2$ of $SU(2)_R$ (index
$i$). It satisfies the following equations
$$
  \partial_\mu G_a^{i\mu} = 0, \qquad G_a^{i\mu} \sigma_\mu^{\dot{a}a}=0
$$
We can construct conserved fermionic currents by multiplying the
supercurrent with a conformal killing spinor $\psi^a(x)$ as
\eqn\cccr{
  j^{i\mu}(x) = \psi^a(x) G_a^{i\mu}(x)}
It is easy to show that 
$$
  \partial_\mu j^{i\mu} = 0
$$
and the corresponding conserved supercharge is $\int d^3x
j^{i0}$. In ${\bf R}^4$ the most general conformal killing spinor
is 
\eqn\cks{
  \psi^a(x) = \lambda^a + x^{\dot{a}a} \mu_{\dot{a}}} where
$\lambda^a$ is an arbitrary constant $(1/2,0)$ spinor, $\mu_{\dot{a}}$
an arbitrary constant $(0,1/2)$ spinor and we use the notation $
x^{\dot{a}a}= x^\mu \sigma_\mu^{ \dot{a}a}$. If we take $\lambda^a\neq
0$ and $\mu_{\dot{a}}=0$ we generate the left chiral
supercharges $Q^i_a$, while for $\lambda^a=0$ and $\mu_{\dot{a}}\neq 0$
we get the right chiral superconformal charges
$\overline{S}^{i\dot{a}}$
$$
Q_a^i \sim \int d^3x\, G^i_{a0},\qquad \overline{S}^{i\dot{a}} \sim \int d^3x\,
x^{\dot{a}a}G^i_{a0}
$$
In a similar way the right chiral supercurrent
$\overline{G}_{i\dot{a}}^\mu$ generates the right chiral supercharges
$\overline{Q}_{i\dot{a}}$ and the left chiral superconformal charges
$S^a_i$.

The OPE of the supercurrent with a scalar primary operator $\varphi$ has
the following form
\eqn\scope{
  G^{i\mu}_a(x) \varphi(0) = ...+{x^\mu x_{a\dot{b}}\over 2\pi^2
|x|^6} [\overline{S}^{i\dot{b}},\varphi](0)+ { x^\mu\over
2\pi^2|x|^4}[Q^i_a,\varphi](0)+...}  For every choice of the
conformal Killing spinor \cks\ we can derive a Ward identity for the
corresponding conserved current $j^{i\mu}$ given in \cccr. We start with
$$
\langle j^{i\mu}(z) \varphi_1(x_1)...\varphi_n(x_n)\rangle
$$
and follow the same steps as before\foot{We also have to use that
$G_a^{i\mu}$ has $\Delta=7/2$, so if it is taken to infinity inside
any correlator, it falls off at least as ${1\over |z|^7}$.}. Being
careful about the expansion of $j^{i\mu}$ around the punctures and
using \scope\ we find the following superconformal Ward identity
\eqn\mainswardl{\eqalign{
  \sum_{k=1}^n \psi^a(x_k) \langle \varphi_1(x_1)&
  ... [Q^i_a,\varphi_k](x_k)...\varphi_n(x_n)\rangle\cr &
  - \sum_{k=1}^n (\partial_\mu \psi^a)\sigma_{a\dot{b}}^\mu
  (x_k) \langle \varphi_1(x_1)
  ... [\overline{S}^{i\dot{b}},\varphi_k](x_k)...\varphi(x_n)\rangle=0}}
  If we take $\psi^a(x)$ to be constant then we find that the usual
  supercharges can be moved around without any factors
\eqn\swic{
 \sum_{k=1}^n \langle
    \varphi_1(x_1)...[Q_a^i,\varphi_k](x_k)...\varphi_n(x_n)\rangle =0}
and if we take $\psi^a(x)$ to be proportional to $x$ we find that for the superconformal partners we have
\eqn\swicd{
 \sum_{k=1}^n \langle \varphi_1(x_1)...[(x_k-x_0)^{\dot{a}a}Q_a^i-\overline{S}^{i,\dot{a}},\varphi_k](x_k)...\varphi_n(x_n)\rangle
    =0} for any $x_0$.

From \mainswardl\ we see that if a field $\varphi$ is a superconformal
primary (hence annihilated by the $\overline{S}$'s) then we can choose
the conformal Killing spinor $\psi^a(x)$ to vanish at the point of
insertion of $\varphi$ and that when applying the superconformal Ward
identity (for this special $\psi^a(x)$) the field $\varphi$ does not
contribute at all, whether it is annihilated by the $Q$'s or not. This
is the essence of identity \basicwardb. In two dimensions it was discussed
in \DixonFJ,\DijkgraafDJ.

For completeness we present the Ward identities for the right chiral
supercurrent $\overline{G}^\mu_{i\dot{a}}$
\eqn\mainswardr{\eqalign{
  \sum_{k=1}^n \psi^{\dot{a}}(x_k) \langle \varphi_1(x_1)&
  ... [\overline{Q}_{i\dot{a}},\varphi_k](x_k)...\varphi_n(x_n)\rangle\cr &
  + \sum_{k=1}^n (\partial_\mu \psi^{\dot{a}})\sigma_{b{\dot{a}}}^\mu
  (x_k) \langle \varphi_1(x_1)
  ... [S^b,\varphi_k](x_k)...\varphi_n(x_n)\rangle=0}}
which implies
\eqn\swie{
    \sum_{k=1}^n \langle \varphi_1(x_1)...[\overline{Q}_{i,\dot{a}},\varphi_k](x_k)...\varphi_n(x_n)\rangle
    =0
}
and
\eqn\swif{
\sum_{k=1}^n \langle
    \varphi_1(x_1)...[(x_k-x_0)^{\dot{a}a}\overline{Q}_{i,\dot{a}}+S_i^{a},\varphi_k](x_k)...\varphi_n(x_n)\rangle
    =0
}
for all $x_0$.

\appendix{B}{Some basic properties of ${\cal N}=2$ superconformal theories}

The supermultiplet containing the stress-energy tensor and the other conserved
currents begins with a scalar superconformal primary operator ${\cal A}$ 
with $\Delta=2,I=0,R=0$. It is a short multiplet of the superconformal 
algebra \SohniusPK.
The other conformal primaries in the multiplet are (diagram is from
Dolan and Osborn \DolanTT)
\eqn\mainmult{
\matrix{\Delta&~~~&{~~~}&&{~~~}&&{~~~}&&{~~~}&&{~~~}&&{~~~}&&\cr
2&&{~~~}&&{~~}&&{\cal A}&&{~~}&&{~~~}&\cr
&&&&&\Bsw&&\Bse&&&&\cr
{{5\over 2}}&&&&\hidewidth~~~~~ \psi_a^i 
\hidewidth&&&&\hidewidth~~\overline{\psi}_{i,\dot{a}}
~~\hidewidth&&&\cr&&&\Bsw&&
\Bse&&\Bsw&&\Bse&&\cr
3&&\hidewidth~~~~~~H_{ab}\hidewidth&&&&
\hidewidth J_\mu, I^i_\mu\hidewidth&&&&
\hidewidth\overline{H}_{\dot{a}\dot{b}}\hidewidth&\cr
&&&\Bse&&\Bsw&&\Bse&&\Bsw&&\cr
{{7\over 2}}&&&
&\hidewidth~~ G_a^{i,\mu} 
\hidewidth&&&
&\hidewidth \overline{G}_{i,\dot{a}}^\mu
\hidewidth&&&\cr
&&&&&\Bse&&\Bsw&&&&\cr 4&&&&& &\hidewidth T_{\mu\nu}\hidewidth&&&&&\cr
R&&~-2&&-1&&0&&1&&2&\cr} } where the $\Bsw$ arrows denote the action
of the left-chiral supercharges $Q_a^i$ and $\Bse$ of
$\overline{Q}_{i,\dot{a}}$.  The conformal dimension $\Delta$ and
$U(1)_R$ charge $R$ of the operators can be seen in the diagram. The
fermions $\psi_a^i$ have spin $(1/2,0)$ and transform in the $I=1/2$
representation of $SU(2)_R$.  The $U(1)_R$ current is $J_\mu$ and the
$SU(2)_R$ currents are $I^i_\mu,i=1,2,3$.  The field $H_{ab}$ is a
spin $(1,0)$ operator singlet under the $SU(2)_R$.  The fields
$G_a^{i,\mu}$ (and their conjugates) are the supercurrents and
transform in the $I=1/2$ of $SU(2)_R$. The stress-energy tensor
$T_{\mu\nu}$ has spin $(1,1)$, $I=0$ and $R=0$.

${\cal N}=2$ SCFTs have two ``central charges'' $c,a$ defined as
coefficients of the trace anomaly on a curved manifold
$$
  \langle T_\mu^\mu\rangle = {c\over 16 \pi^2} ({\rm Weyl})^2 -
  {a \over 16 \pi^2} ({\rm Euler})
$$
The central charge $c$ is related to the 2-point function of the stress-energy tensor as
%% The 2-point functions of fields in the supermultiplet of ${\cal A}$
%% have the form
%% \eqn\curtwop{\eqalign{
%% &\langle{\cal A}(x) {\cal A}(0)\rangle = {3c\over 8 \pi^4}{1\over
%% |x|^4}\cr &\langle J_\mu(x) J_\nu(0) \rangle = {12 c \over \pi^4}
%% {I_{\mu\nu}(x)
%% \over |x|^6}\cr
\eqn\curtwop{
 \langle T_{\mu\nu}(x) T_{\rho\sigma}(y)\rangle = {40 c \over \pi^4}
{{\cal I}_{\mu\nu,\rho\sigma}(x)\over |x|^8} } where
$$
 I_{\mu\nu}(x) \equiv g_{\mu\nu} - 2 {x_\mu x_\nu \over |x|^2}
$$
$$
{\cal I}_{\mu\nu,\rho\sigma} = {1\over
2}\left(I_{\mu\sigma}(x)I_{\nu\rho}(x) +I_{\mu\rho}(x)
I_{\nu\sigma}(x) - {1\over 4}\delta_{\mu\nu}\delta_{\rho\sigma}\right)
$$

\appendix{C}{Contours}

Now let us consider the following quantity
\eqn\contoursh{\eqalign{
  \con= &{1\over (2\pi)^4}\lim_{r\rightarrow 1^-}\int_{|x|=r}d\Omega_3^x \int_{|y|=1}
 d\Omega_3^y |x|^2|y|^2 (y\cdot \partial_y)(x\cdot \partial_x) \cr
 & \left( { |y|^2\over
 |x|^2} \langle \overline{\bphi}_l(\infty) \lphi_i(x) \overline{\lphi}_j(y) \bphi_k(0)\rangle
 -{|x|^2 \over
 |y|^2} \langle \overline{\bphi}_l(\infty) \lphi_i(y) \overline{\lphi}_j(x) \bphi_k(0)\rangle
\right) }}
In this double integral the variable $y$ lies on a 3-sphere of unit
radius while $x$ on a 3-sphere of radius $r$ and we are taking the
limit $r\rightarrow 1^-$. So let us parametrize
$$
  y =1\cdot \hat{\Omega}_3^y,\qquad x = r\cdot \hat{\Omega}_3^x
$$
If $\hat{\Omega}_3^y \neq \hat{\Omega}_3^x$ then the contributions
between the two points $(\hat{\Omega}_3^y, \hat{\Omega}_3^x)$ and
$(\hat{\Omega}_3^x,\hat{\Omega}_3^y)$ cancel in the limit
$r\rightarrow 1$.  So one might conclude that the integral vanishes,
however the previous argument cannot be applied for the point
$\hat{\Omega}_3^y = \hat{\Omega}_3^x$ where we may have a
$\delta$-function-like contribution. To check the contribution to $\con$
from the region where $x\rightarrow y$ we can use the OPE between
$\lphi_i$ and $\overline{\lphi}_j$. 

In order to include  at once the contribution of all descendants for each
primary, it is more convenient to perform the expansion in
conformal partial waves in the $(i\overline{j})\rightarrow
(k\overline{l})$ channel. This means that we can write the 4-point function as
$$
\langle \overline{\phi}_l(\infty)\Phi_i(x)\overline{\Phi}_j(y) \phi_k(0)\rangle
= {1\over |x-y|^4} \sum_{\cal O} C_{i\overline{j}}^{\cal O} C_{{\cal O} k \overline{l}} \,\,g_{\Delta,l}(u,v)
$$
where the sum is over all conformal primary (though not necessarily
superconformal primary) operators ${\cal O}$ of dimension $\Delta$ and
spin\foot{In the OPE of two scalars we do not have any primary
operators of spin $(j,\overline{j})$ with $j\neq \overline{j}$. Here
we take $l=j+\overline{j}$.} $l$. The quantities $u,v$ are the
conformal cross-ratios which for our configuration of 4-points
$x,y,0,\infty$ are
\eqn\ccross{
u = {|x-y|^2 \over |x|^2},\qquad v = {|y|^2\over |x|^2}
}
The function $g_{\Delta,l}(u,v)$ is the conformal partial
wave. Explicit expressions for these functions can be found in \DolanUT.

The other 4-point function $\langle \overline{\phi}_l(\infty)\Phi_i(x)\overline{\Phi}_j(y) \phi_k(0)\rangle$ can be expanded similarly with the roles of $x,y$ interchanged which is equivalent to the substitution $u\rightarrow u/v,\, v\rightarrow 1/v$. So we have
$$
\langle \overline{\phi}_l(\infty)\Phi_i(y)\overline{\Phi}_j(x) \phi_k(0)\rangle
= {1\over |x-y|^4} \sum_{\cal O} C_{i\overline{j}}^{\cal O} C_{{\cal O} k \overline{l}} \,\,g_{\Delta,l}(u/v,1/v)
$$
Now we use a basic property of the conformal partial waves \DolanUT\
$$
g_{\Delta,l}(u/v,1/v) = (-1)^l g_{\Delta,l}(u,v)
$$
We can thus finally write the integral that we want to compute as
$$
{\cal E} = \sum_{\cal O} C_{i\overline{j}}^{\cal O} C_{{\cal O}k\overline{l}}\,\, X_{\Delta,l}
$$
where
$$
X_{\Delta,l} \equiv {1\over (2\pi)^4}\lim_{r\rightarrow 1^-}\int_{|x|=r}d\Omega_3^x \int_{|y|=1}
 d\Omega_3^y |x|^2|y|^2 (y\cdot \partial_y)(x\cdot \partial_x)  \left( { |y|^4 -(-1)^l |x|^4 \over |x|^2|y|^2} g_{\Delta,l}(u,v)\right)
$$
and in this integral the parameters $u,v$ have to be computed from \ccross. Notice
that the quantities $X_{\Delta,l}$ are independent of dynamical information of the CFT and only depend on kinematics of the conformal group.

After some work one can verify that the only nonzero terms are the following
\eqn\nonzerocont{\eqalign{
& X_{0,0} = 1\cr & X_{2,0} = 2 \cr & X_{3,1} =1 }} The first term corresponds to the identity
operator. The second from the conformal partial wave of the special
superconformal primary ${\cal A}$ of $\Delta=2,l=0$ whose
supermultiplet contains the stress energy tensor, as explained in the
previous appendix\foot{We assume that this operator is unique,
otherwise there would be two conserved spin-2 currents. This
assumption is not correct in the free limit of ${\cal N}=2$
superconformal gauge theories where we also have the Konishi scalar
with $\Delta=2$, in addition to ${\cal A}$. However, when we turn on
the coupling the Konishi multiplet gets anomalous dimensions. So we
will assume that a generic point on the moduli space of the CFT there
is no other scalar with the same quantum numbers as ${\cal A}$. It
would be interesting to explore this assumption in more detail.}.  The
third term corresponds to the $U(1)_R$ current\foot{If we want the marginal operator to be neutral under the
current (otherwise the symmetry would not be preserved after the
marginal deformation and it would be ``accidental'' and not present at
a generic point on the moduli space), then only the $U(1)_R$ current
can appear, since the supercharges carry R-charge and can make the
marginal operator neutral even if the chiral primary itself is charged
( supercharges are uncharged under the non R-symmetries).}. So all in all we find
$$
{\cal E} = g_{i\overline{j}} g_{k\overline{l}} + 2\, C_{i\overline{j}}^{\cal A} C_{{\cal A}k\overline{l}}+ C_{i\overline{j}}^J C_{J k \overline{l}}
$$
What is missing now is to calculate the quantities $
C_{i\overline{j}}^{\cal A} C_{{\cal A}k\overline{l}}$ and
$C_{i\overline{j}}^J C_{J k \overline{l}}$. To proceed we will use the
following logic. The Ward identities fix that the conformal
partial wave of the stress tensor is weighed by the overall coefficient
$C_{i\overline{j}}^T C_{Tk\overline{l}} = {\Delta_i \Delta_k \over 90
c} g_{i\overline{j}}g_{k\overline{l}}$ (see \DolanUT). Using
$\Delta_i=2$ and $\Delta_k=R/2$ we find
$$
 C_{i\overline{j}}^T C_{Tk\overline{l}}  = {R\over 90c} g_{i\overline{j}}g_{k\overline{l}}
$$
In theories with ${\cal N}=1$ and ${\cal N}=2$ superconformal invariance, supersymmetry relates the weight of the conformal partial wave of the stress tensor to the weights of the partial waves of the other members of the superconformal multiplet containing the stress tensor. This can be found in \DolanTT\ and more recently in \PolandWG. Using their results we find for the $U(1)_R$ current
$$
C_{i\overline{j}}^J C_{Jk \overline{l}} = -{15 \over 2} C_{i\overline{j}}^T C_{Tk\overline{l}} = 
-{R\over 12c} g_{i\overline{j}}g_{k\overline{l}}
$$
and for the operator ${\cal A}$
$$
C_{i\overline{j}}^{\cal A}C_{{\cal A}k\overline{l}} = 15  C_{i\overline{j}}^T C_{Tk\overline{l}}
= {R\over 6c} g_{i\overline{j}}g_{k\overline{l}}
$$
These relations are explained in some more detail in the next appendix. Putting everything together we find
$$
{\cal E} =  \left(1 +{R\over 4c} \right) g_{i\overline{j}}g_{k\overline{l}}
$$

\appendix{D}{Conformal block decomposition and supersymmetry}

Let us explain how we fix the relative normalization of the conformal partial waves
in ${\cal N}=2$ superconformal field theories. We first consider the ${\cal N}=2$ theory as an ${\cal N}=1$ superconformal field theory and then we use the results of \PolandWG. An ${\cal N}=1$ SCFT has a $U(1)_R$ R-symmetry. The corresponding current $r^\mu$ is related to the ${\cal N}=2$ R-currents by
\eqn\rcura{
r^\mu = {1\over 3} J^\mu +{4\over 3} I_3^\mu
}
where $J^\mu$ is the ${\cal N}=2$ $U(1)_R$ current and $I_3^\mu$ one of the $SU(2)_R$ currents of the ${\cal N}=2$ theory. The orthogonal linear combination of these two
currents
\eqn\rcurab{
f^\mu =J^\mu -2  I_3^\mu
}
plays the role of a flavor current from the ${\cal N}=1$ point of view. The overall
normalization of $f$ is not important for us since it will drop out, 
while the relative coefficients were fixed by requiring the orthogonality\foot{i.e. that the 2-point function$\langle r^\mu f^\nu\rangle =0$. Notice that the currents $J_\mu,I_3^\mu$ are orthogonal.} of the currents $r^\mu,f^\mu$. Since the chiral primaries are only charged under the current $J_\mu$ (and not $I_3^\mu$) we have
\eqn\sumcurr{
C_{i\overline{j}}^J C_{Jk\overline{l}}=C_{i\overline{j}}^r C_{rk\overline{l}}+C_{i\overline{j}}^f C_{fk\overline{l}}
}
On the other hand from \rcura, \rcurab\ we have
$$
C_{i\overline{j}}^r C_{rk\overline{l}} = {1\over 9} C_{i\overline{j}}^J C_{Jk\overline{l}} \times {
g_{JJ} \over g_{JJ}/9 + 16 g_{II}/9}
$$
$$
C_{i\overline{j}}^f C_{fk\overline{l}} =  C_{i\overline{j}}^J C_{Jk\overline{l}} \times {g_{JJ} \over  g_{JJ} +  4 g_{II}}
$$
where by $g_{JJ},g_{II}$ we denote the 2-point function of the
corresponding current.  The ${\cal N}=2$ algebra fixes $g_{JJ}=8\,
g_{II}$ (see for example \ArgyresCN) and we find
\eqn\ratiocur{
{C_{i\overline{j}}^r C_{rk\overline{l}} \over C_{i\overline{j}}^f C_{fk\overline{l}} } = {1\over 2}
}
Combining \sumcurr\ with \ratiocur\ we find that 
\eqn\curcur{
C_{i\overline{j}}^J C_{Jk\overline{l}}=3 \,C_{i\overline{j}}^r C_{rk\overline{l}} = {3\over 2}
\,C_{i\overline{j}}^f C_{fk\overline{l}}}

In \PolandWG\ the relative weight of conformal partial waves of operators
in the same superconformal multiplet in ${\cal N}=1$ SCFTs were presented. The ${\cal  N}=1$ R-current $r^\mu$ is in the same ${\cal N}=1$ superconformal multiplet as the stress tensor. From their results we find that
$
C_{i\overline{j}}^r C_{rk\overline{l}} = -{5\over 2} C_{i\overline{j}}^T C_{Tk\overline{l}}
$
and using \curcur\ we have
\eqn\finaleqa{
C_{i\overline{j}}^J C_{Jk\overline{l}} = -{15 \over 2}
C_{i\overline{j}}^T C_{Tk\overline{l}} } 
Similarly, the flavor current
$f^\mu$ is in the same multiplet as the scalar ${\cal A}$ of
$\Delta=2$,$R=0$. From the results of \PolandWG\ we have
$
C_{i\overline{j}}^{\cal A}C_{{\cal A}k\overline{l}} = -3 C_{i\overline{j}}^f C_{fk\overline{l}}
$
and using \curcur, \finaleqa\ we find
\eqn\finaleqb{
C_{i\overline{j}}^{\cal A}C_{{\cal A}k \overline{l}} = 15\, C_{i\overline{j}}^T C_{Tk\overline{l}}
}
The relations \finaleqa\ and \finaleqb\ are the ones that we need.

\appendix{E}{Curvature of supercurrents}

Here we show that the curvature of supercurrents is given by expressions \ccur.
The curvature of the supercurrents $F_{\mu\nu}^{\cal L}$ can be computed as the relative curvature between the chiral primaries $\phi_k$ and their first descendants $Q_a^I\phi_k$, since we have\foot{We supress the indices $I,a$ in this formula since the
curvature is diagonal in $I,a$.}
$$
(\widetilde{F}_{\mu\nu})_k^l = (F_{\mu\nu})_k^l + F_{\mu\nu}^{\cal L}\, \delta_k^l
$$
where $(F_{\mu\nu})_k^l$ is the curvature of the chiral primaries computed in \maintt\ and $(\widetilde{F}_{\mu\nu})_k^l$ is the curvature of the descendants $Q_a^I$ that we will try to compute now. To do it we consider the formula \maincurv\ for operators of the form $Q_a^I \phi_k$. 

First let us show that the line bundle ${\cal L}$ of the supercurrents
is holomorphic, i.e. that $F_{ij}^{\cal L} = F^{\cal
L}_{\overline{i}\, \overline{j}} = 0$.  In \maintt\ we showed that $(F_{ij})_k^l=0$ so we simply have
to show that $(\widetilde{F}_{ij})_k^l$=0. For this it is sufficient to show
the vanishing of the 4-point function
$$
A=\langle (\overline{Q}_{I,\dot{b}} \overline{\phi}_l)(z) {\cal O}_i(x) {\cal O}_j (y) (Q_a^I \phi_k)(0)\rangle
$$
Notice that there is no summation over $I$ implied. In the rest of this appendix we simply choose $I$ to be either 1 or 2, the answer does not depend on this choice. %% It is convenient to first consider
%% $$
%% \widetilde{A}(z) = \langle (\overline{Q}_{I,\dot{b}} \overline{\phi}_l)(z) {\cal O}_i(x) {\cal O}_j (y) (Q_a^I \phi_k)(0)\rangle
%% $$
Using the Ward identity \swic\ we can move the supercharge $Q_a^I$ away from the point $0$. The supercharge $Q_a^I$ annihilates the operators ${\cal O}_i,{\cal O}_j,\overline{\phi}_l$, so using the algebra we find
$$
A = 2 \partial^z_{a\dot{b}} \,
\langle \overline{\phi}_l(z) {\cal O}_i(x) {\cal O}_j (y) \phi_k(0)\rangle=0
$$
because, as we explained in equation \holhol, the 4-point function appearing above vanishes. From \maincurv\ it follows that $(\widetilde{F}_{ij})_k^l=0$ and we find 
$$
F_{ij}^{\cal L} = 0
$$
Similarly we show that $F_{\overline{i}\,\overline{j}}^{\cal L} = 0$. Thus we have shown that ${\cal L}$ is a holomorphic line bundle over the moduli space.

Now let us compute the nonzero part of the curvature $F_{i\overline{j}}^{\cal L}$. First we evaluate
the 2-point function of descendant operators. If we have
$$
\langle \overline{\phi}_l(x) \phi_k(y) \rangle ={g_{k\overline{l}}\over |x-y|^{2\Delta}}
$$
then using the Ward identities we find
$$
\langle (\overline{Q}_{I,\dot{b}} \overline{\phi}_l) (x) (Q^I_a \phi_k)(y) \rangle =-4\Delta  g_{k\overline{l}}{(x-y)_{a\dot{b}}\over |x-y|^{2\Delta+2}}
$$
The correlator relevant for the computation of the curvature of the descendants $Q_a^I \phi_k$ is
$$
B=-\lim_{z\rightarrow \infty}\left( {z^{\dot{b}a} |z|^{2\Delta}\over 8\Delta}  g^{
\overline{m}l} \langle (\overline{Q}_{I,\dot{b}} \overline{\phi}_m)(z) {\cal O}_i(x) \overline{\cal O}_j (y) (Q_a^I \phi_k)(0)\rangle\right)
$$
and a similar one with $x\leftrightarrow y$. Here we have multiplied the 4-point function with the inverse 2-point function of the descendant at infinity in order to raise the indices. We have also taken a trace over the spinor indices to simplify intermediate equations\foot{Here we are taking advantage of the fact that the curvature of operators $Q_a^I\phi_k$ is the same for all $a$.} and divided by 2 to keep the normalization correct. Using the superconformal Ward identity \swicd\ to move $Q_a^I$ away from $0$ we find that
\eqn\betaf{\eqalign{
& B=  g^{\overline{m}l} \langle \overline{\phi}_m(\infty) {\cal O}_i(x) \overline{\cal O}_j(y) \phi_k(0)\rangle \cr 
&  -\lim_{z\rightarrow \infty}\left( {(y-z)^{\dot{b}a} |z|^{2\Delta}\over 8\Delta}  g^{
\overline{m}l} \langle (\overline{Q}_{I,\dot{b}} \overline{\phi}_m)(z) {\cal O}_i(x) (Q_a^I \overline{\cal O}_j) (y) \phi_k(0)\rangle\right) 
\cr
& +\lim_{z\rightarrow \infty}\left( { |z|^{2\Delta}\over 8\Delta}  g^{
\overline{m}l} \langle (\overline{Q}_{I,\dot{b}} \overline{\phi}_m)(z) {\cal O}_i(x) (\overline{S}^{\dot{b},I} \overline{\cal O}_j) (y) \phi_k(0)\rangle\right) 
}}
For the first term we used that for antichiral primaries $\{\overline{S}^{\dot{b},I},[ \overline{Q}_{I,\dot{b}} ,\overline{\phi}_m]\} = -8 \Delta \overline{\phi}_m$ (as we explained above, here we are summing over $\dot{b}$ but not $I$). After plugging the expression for $B$ (and the corresponding one for $x\leftrightarrow y$)  into \maincurv, the first term in \betaf\ gives the curvature $(F_{i\overline{j}})_k^l$ of the superconformal primaries $\phi_k$. The third term in \betaf\ is actually zero in the limit $z\rightarrow \infty$. Only the second term in \betaf\ is relevant that we define as
$$
C = \lim_{z\rightarrow \infty}\left( {z^{\dot{b}a} |z|^{2\Delta}\over 8\Delta}  g^{
\overline{m}l} \langle (\overline{Q}_{I,\dot{b}} \overline{\phi}_m)(z) {\cal O}_i(x) (Q_a^I \overline{\cal O}_j) (y) \phi_k(0)\rangle\right) 
$$
where we dropped a factor of $y$ since it does not contribute in the limit $z\rightarrow \infty$. Similarly we define $\widetilde{C}$ with $x\leftrightarrow y$.

So the curvature of the supercurrents is given by
\eqn\supercurrentcurv{
F_{i\overline{j}}^{\cal L} \, \delta_k^l = {1\over (2\pi)^4} \int_{|x|\leq 1} d^4 x \int_{|y|\leq 1} d^4 y \left(C-\widetilde{C}\right)
}
where the integrals have to be regularized in the way described in the main text. Using the superconformal Ward identity  \swif\ to move $\overline{Q}_{I,\dot{b}}$ away from $z$ we find that
\eqn\asdfasdf{\eqalign{
C = & -{g^{\overline{m}l}\over 8 \Delta} \Big(y^{\dot{b}a} \langle  \overline{\phi}_m(\infty)
{\cal O}_i(x) \overline{Q}_{I,\dot{b}}Q_a^I \overline{\cal O}_j(y) \phi_k(0)\rangle  + 
\langle  \overline{\phi}_m(\infty)
{\cal O}_i(x) S_I^a Q_a^I \overline{\cal O}_j(y) \phi_k(0)\rangle \cr
&
+x^{\dot{b}a}  \langle  \overline{\phi}_m(\infty)
\overline{Q}_{I,\dot{b}}{\cal O}_i(x) Q_a^I \overline{\cal O}_j(y) \phi_k(0)\rangle
+  \langle  \overline{\phi}_m(\infty)
S_I^a{\cal O}_i(x) Q_a^I \overline{\cal O}_j(y) \phi_k(0)\rangle\Big)}}
Massaging this expression a little more we find
\eqn\finalwardsu{\eqalign{
 C ={ g^{\overline{m}l}\over 8 \Delta}&\Big[ \Big( 4(x-y)\cdot \partial_y -16 \Big) \Boxx\Boxx \langle  \overline{\phi}_m(\infty)
\phi_i(x) \overline{\phi}_j(y) \phi_k(0)\rangle\cr &
+ 16 \Boxx (\partial_x\cdot \partial_y) \langle  \overline{\phi}_m(\infty)
\phi_i(x) \overline{\phi}_j(y) \phi_k(0)\rangle\Big]
}}
and a similar answer for $\widetilde{C}$ with the substitution $x\leftrightarrow y$.

One can check that in the integration over $x,y$ in \supercurrentcurv\ then the expression \finalwardsu\ (and the one for $\widetilde{C}$) gives a finite contribution from the region $y\rightarrow 0 $ and $x\rightarrow 0$. Hence to evaluate \supercurrentcurv\ one can integrate by parts and simply pick up the terms from the hemispheres $y\rightarrow 1$ and $x\rightarrow 1^-$ as we did in section 5 and appendix C.

The partial integration leads to the following expression
\eqn\fdsfdsf{\eqalign{
F_{i\overline{j}}^{\cal L} =&  {g^{\overline{m}l} \over 8\Delta}{1\over (2\pi)^4} \lim_{r\rightarrow 1^-} \int_{|x|=r} d\Omega_3^x
\int_{|y|=1}d\Omega_3^y |x|^2|y|^2
\cr &
  \Big\{(y\cdot\partial_y)(x\cdot \partial_x)\left[\left(4(x-y)\cdot\partial_y -8\right){|y|^2\over |x|^2} \langle  \overline{\phi}_m(\infty)
\phi_i(x) \overline{\phi}_j(y) \phi_k(0)\rangle  - (x\leftrightarrow y) \right]
\cr &
+ 8
\left[(y\cdot \partial_y)(x\cdot\partial_y)  {|y|^2\over |x|^2} \langle  \overline{\phi}_m(\infty)
\phi_i(x) \overline{\phi}_j(y) \phi_k(0)\rangle  - (x\leftrightarrow y) \right] \Big\}
}}
As we explained in appendix C, we can evaluate these integrals by considering the conformal partial wave expansion of the 4-point function in the channel $(i\overline{j})\rightarrow (k\overline{l})$. Following the analogous steps as in appendix C we find that in this expansion only the conformal partial wave of the operator ${\cal A}$ and of the $U(1)_R$ current $J$ contribute to these integrals and that
$$
 F_{i\overline{j}}^{\cal L} 	=-12 {C_{i\overline{j}}^{\cal A} C_{{\cal A}k\overline{l}} +  C_{i\overline{j}}^{\cal J} C_{{\cal J}k\overline{l}}\over 8\Delta}
$$
using $\Delta = R/2$ and the results from the previous appendices we finally find
$$
 F_{i\overline{j}}^{\cal L} = -{1\over 4c} g_{i\overline{j}}
$$

\appendix{F}{Chiral-Antichiral OPE}

Let us consider the OPE of a scalar antichiral primary
$\overline{\Phi}_j$ of $\Delta=2$, $R=-4$ with a (for simplicity)
scalar chiral primary $\phi_k$ of $\Delta>2$ and $R=2\Delta$. Consider
a scalar primary operator ${\cal O}$ appearing on the RHS with
dimension $\Delta'$ and $U(1)_R$ charge $R'$. We have
\eqn\cacope{
\overline{\Phi}_j(x) \phi_k(0) \sim D_{\overline{j}k}^{\cal O} {{\cal O}(0) \over |x|^{2+\Delta -\Delta'}}+...}
From $U(1)_R$ charge conservation we have $R' = 2\Delta -4$ and
unitarity implies $\Delta' \geq {R' \over 2}$ or $\Delta' \geq \Delta
- 2$. We find that the term on the RHS of \cacope\ goes like ${1\over
|x|^4}$ when this inequality is saturated $\Delta' = \Delta-2$ which
implies that ${\cal O}$ is one of the chiral primaries, say
$\phi_n$. Then it is not difficult to show that
$$
D_{\overline{j}k}^{\cal O} = g_{k\overline{r}}C^{*\overline{r}}_{
\overline{j}\overline{p}} g^{\overline{p}n}
$$
\listrefs 

\end